\documentclass[]{interact}

\usepackage{epstopdf}% To incorporate .eps illustrations using PDFLaTeX, etc.
\usepackage[caption=false]{subfig}% Support for small, `sub' figures and tables

\usepackage[numbers,sort&compress]{natbib}% Citation support using natbib.sty
\usepackage{latexsym}
\usepackage{stackrel}
\usepackage{mathtools}
\usepackage{multirow}
\usepackage{amsmath}
\usepackage[makeroom]{cancel}
\usepackage{rotating}
\usepackage{pdflscape}
\usepackage[T1]{fontenc}  % access \textquotedbl
\usepackage{textcomp}     % access \textquotesingle
\usepackage{hyperref}

\bibpunct[, ]{[}{]}{,}{n}{,}{,}% Citation support using natbib.sty
\renewcommand\bibfont{\fontsize{10}{12}\selectfont}% Bibliography support using natbib.sty
\makeatletter% @ becomes a letter
\def\NAT@def@citea{\def\@citea{\NAT@separator}}% Suppress spaces between citations using natbib.sty
\makeatother% @ becomes a symbol again

\theoremstyle{plain}% Theorem-like structures provided by amsthm.sty

\theoremstyle{definition}

\theoremstyle{remark}

\begin{document}

%\articletype{ARTICLE TEMPLATE}% Specify the article type or omit as appropriate

\title{Assessment of subgrid scale mixing models used in LES at high pressures}
\thanks{This is an Accepted Manuscript of an article published by Taylor \& Francis in \textit{Journal of Turbulence} on 17 July 2018. The Version of Record is: Neelakantan Padmanabhan and Richard S. Miller (2018), ``Assessment of subgrid scale mixing models used in LES at high pressures,'' \textit{Journal of Turbulence}, 19(8), 683--715. https://doi.org/10.1080/14685248.2018.1498590}
\author{
\name{Neelakantan Padmanabhan \thanks{CONTACT N.Padmanabhan Email: neel.p.3187@gmail.com} and Richard S. Miller}
\affil{Department of Mechanical Engineering, Clemson University, Clemson, South Carolina 29634}
}

\maketitle

\begin{abstract}
Molecular mixing models for the conditional scalar diffusion (CSD) term in the filtered mass density function (FMDF) transport equation used in Large Eddy Simulation (LES-FMDF) approach are evaluated against direct numerical simulation (DNS) of spatially evolving jets. A database of DNS of turbulent mixing slot jets of $C_{7}H_{16} / O_2$ and $C_{7}H_{16} / Air$ is developed. The formulation includes the compressible form of the governing equations, a generalized multicomponent diffusion model, a cubic real gas equation of state and realistic property models. Simulations are conducted over a wide range of initial pressures ($1 atm < P_0 < 100 atm$) and jet width based Reynolds numbers of 850 and 1300. FMDF of mixture fraction at various filter widths are obtained from the simulation at several spatial locations within the flow. The CSD term in the exact transport equation of FMDF is calculated from the DNS. An \textit{a priori} analysis of Interaction by Exchange of Mean (IEM), Modified Curl (MC) and Mapping Closure (MAPPING) mixing models for CSD is conducted. At high pressures, significant deviations are observed between the CSD measured from the DNS and those predicted by the models. The source of the deviations is determined to be the mixing frequency used in the models. The significance of the mixing frequency and its variation with change in pressure and diffusion models is studied. The primary objective of this work is to assess the applicability of aforementioned mixing models to flows at large pressures and determine the causes for deviation in prediction.
\end{abstract}
%%%%%%%%%%%%%
\section{Introduction}
\label{intro}
Many modern day combustion devices such as diesel engines and gas turbines operate at pressures higher than atmospheric pressure ($P \ge 10 atm$), while rocket engines operate at pressures excess of $100 atm$  \cite{paperJ}. Effective design and development of such devices requires simulation tools that employ Reynolds Averaged Navier Stokes (RANS) and Large Eddy Simulation (LES) approaches which rely on accurate models to reproduce the effects of turbulent mixing and combustion. Traditional LES models may be broadly classified as functional and structural models which typically aggregate the effects of small scale kinetic energy diffusion and dissipation into a subgrid scale term. The dissipation is assumed to take place at the scale determined by the filter size for LES, instead of the Kolmogorov scale. A detailed survey of turbulent combustion models for LES of propulsive flow fields is presented in \cite{lessurvey}. These models reproduce the effects of turbulent mixing satisfactorily but result in large errors in capturing the effects of chemical reaction since the chemical source term is highly non-linear. A major relief in this area was obtained after introduction of the Probability Density Function (PDF) method by Pope \cite{popereac}, the Filtered Density Function (FDF) method by Colucci \textit{et.al.} \cite{fdfintro} and Filtered Mass Density Function (FMDF) by Jaberi \textit{et.al.} \cite{fmdfintro}. In these methods, the complete statistical information of the flow is provided by a joint PDF or a filtered joint PDF (in case of FDF and FMDF) for velocity and scalars of the flow. The moments of statistics of the flow are obtained from the solutions of a transport equation derived for the PDF.

The main advantage of this approach is that the chemical source term is closed and does not require modelling. However, this transport equation results in some other terms that remain unclosed and require modelling. Two such terms that arise from this equation are subgrid scale (SGS) convection and conditional scalar diffusion (CSD), which physically represent the effects of molecular convection and molecular diffusion, respectively. The SGS convection is usually closed using a functional model like Smagorinsky or Eddy Viscosity model. The CSD is closed using models like Interaction by Exchange of Mean (IEM), Modified Curl (MC), Mapping Closure (MAPPING), and Euclidean Minimum Spanning Tree (EMST). While some models are more applicable than others in certain scenarios, each of the aforementioned models represent the molecular mixing phenomena fairly well. However, these models were predominantly developed using experimental data and direct numerical simulations (DNS) performed at $1 atm$ pressure with ideal gas assumptions. These models have been observed to perform well at atmospheric pressures. The efficacy of these models at high pressures is yet unknown.

This paper investigates the applicability of a few molecular mixing models used in LES, in flows at high pressures, where multicomponent differential and cross diffusion effects may become much more significant. The models are evaluated with a database of DNS of spatially evolving turbulent slot jets at pressures ranging from $1 atm$ to $100 atm$ with real gas effects, realistic property models and a generalized diffusion model. The spatially evolving turbulent jet enables measurement of FMDF and CSD from the flow at various spatial locations.
%%%%%%%%%%%%%%%%%%%%%%%%%%%%%%%
\section{DNS of a turbulent slot jet}
The DNS solves the fully compressible form of continuity, momentum, total energy and species mass fraction equations. This work incorporates both ideal gas and real gas equations of state, realistic property models, and multicomponent diffusion.
% equations
\begin{equation} \label{cont}
    \frac{\partial \rho}{\partial t}+\frac{\partial}{\partial x_j} [\rho u_j] = 0,
\end{equation}
\begin{equation} \label{NS}
    \frac{\partial}{\partial t}(\rho u_i) + \frac{\partial}{\partial x_j} [\rho u_i u_j + P \delta_{ij} - \tau_{ij}] = 0,
\end{equation}
\begin{equation} \label{energy}
    \frac{\partial}{\partial t} (\rho e_t) + \frac{\partial}{\partial x_j} [(\rho e_t + P)u_j - u_i \tau_{ij} + Q_j + \sum^{N}_{\beta = 1} \overline H^\beta,\overline J^\beta_j] = 0,
\end{equation}
\begin{equation} \label{species}
    \frac{\partial}{\partial t} (\rho Y^\beta) + \frac{\partial}{\partial x_j} [\rho Y^\beta u_j + J^\beta _j] = 0.
\end{equation}
%-------------------------------
In the above equations $t$ represents time, $x_j$ the spatial vector, $\rho$ the density of mixture, $u_j$ the mixture velocity vector, $P$ the total pressure, $\delta_{ij}$ the Kronecker delta tensor, $\tau_{ij}$ the viscous stress tensor, $e_t$ the total specific energy (internal energy plus kinetic energy), $Q_j$ the heat flux vector, $\sum^{N}_{\beta = 1} \overline H^\beta,\overline J^\beta_j$ the enthalpy flux with N species, $\overline H^\beta = \partial \overline H/\partial X_{\beta}$ the partial molar enthalpy of species $\beta$, $X_{\beta}$ the mole fraction of $\beta$, $\overline J^\beta _j$ the molar mass flux vector for species $\beta$ ($J^\beta _j = M_{\beta} \overline J^\beta _j$, where $M_{\beta}$ represents the molecular weight of species $\beta$), and $Y^\beta$ the mass fraction of species $\beta$.

Real gas effects for high pressure simulations are included via the cubic Peng Robinson equation of state. The Peng Robinson equation of state is known to be computationally efficient and easy to implement \cite{pallePhD}. It is expressed as,
\begin{equation}
    P = \frac{\overline{R}T}{\overline{v}-B_m} - \frac{A_m}{\overline{v}^2 + 2 \overline{v}B_m - B_m ^2},
\end{equation}
where $\overline R$ represents the universal gas constant, $T$ the temperature, $\overline v$ the molar volume and $A_m, B_m$ the mixture parameters defined for the Peng Robinson equation of state \cite{pallePhD,vasudevan}. Ideal gas effects for simulations at atmospheric pressures are included via ideal gas equation of state \cite{aysePhD},
\begin{equation}
    P = \rho \sum_{\beta} \frac{\overline R}{M_{\beta}} T.
\end{equation}
The heat flux and mass flux vectors with multicomponent differential and cross diffusion effects applicable to high pressure simulations are derived from Non-Equilibrium Thermodynamics (NEQT) and Keizer's fluctuation theory. The current work only provides the form of the molar mass flux vector. Details of the heat flux vector may be found in \cite{palle,vasudevan,paperJ}
\begin{equation} \label{mass flux}
\begin{split}
    \overline{J}_j ^{\beta} = -n D_m ^{\beta \gamma} \sum_{\gamma \neq \beta} ^N \left\{X_{\beta} X_{\gamma} \frac{M_{\gamma}}{M_m} \alpha_{BK} ^{\beta \gamma} \right\} \frac{\partial lnT}{\partial x_j} \\
    - \sum_{\gamma \neq \beta} ^N \frac{n D_m ^{\beta \gamma}}{\overline{R}T} \left\{- \frac{M_{\beta} M_{\gamma}}{M_m M_m} X_{\beta} X_{\gamma} \overline{v_{,\gamma}} + \frac{M_{\gamma} M_{\gamma}}{M_m M_m} X_{\beta} X_{\gamma} \overline{v_{,\beta}} \right\} \frac{\partial P}{\partial x_j} \\
    - \sum_{\eta=1} ^{N-1} \left\{ \sum_{\gamma \neq \beta} ^N \Big[- \frac{M_{\beta} M_{\gamma}}{M_m M_m} X_{\beta} n D_ m ^{\beta \gamma} \alpha_D ^{\gamma \eta} + \frac{M_{\gamma} M_{\gamma}}{M_m M_m} X_{\gamma} n D_ m ^{\gamma \beta} \alpha_D ^{\beta \eta}  \Big] \right\} \frac{\partial X_{\eta}}{\partial x_j}.
\end{split}
\end{equation}
The general forms of the heat and mass flux vectors were first derived by Harstad and Bellan \cite{harstadbellan} for binary species mixing and extended to arbitrary number of species by Palle \cite{pallePhD}. In the above equations, $n$ represents the molar density ($n = \rho/M_m$), $M_m=\sum_{\beta=1} ^N X_{\beta} M_{\beta}$ the molecular weight of the mixture, $\overline{v_{,\beta}}$ the partial molar volume of species $\beta$, $D_m ^{\beta \gamma}$ the mass diffusivities for diffusion of species $\beta$ into species $\gamma$, $\kappa$ the mixture thermal conductivity, and $\alpha_{BK} ^{\beta \gamma}$, $\alpha_D ^{\beta \eta}$ the thermal and mass diffusion factors. The property models are based on principles of corresponding states or experimental data. Mixture viscosity and thermal conductivity are calculated by Lucas method \cite{propgl} and the method of Steil and Thodos \cite{propgl}, respectively. Heat capacities are derived directly from the equation of state. The binary diffusion coefficients for low pressures are obtained by the method of Fuller \textit{et al.} \cite{propgl}. Coefficients for high pressures are obtained from the correlations of Takahashi \cite{propgl}. Other significant differences between experimental data and property models are corrected by curve fitting models developed by Palle \cite{pallePhD}. Binary thermal diffusion factors $\alpha_{BK} ^{\beta \gamma}$ for high pressures are obtained from curve fitting of experimental data by Vasudevan \cite{vasudevan}. 

For $'Le=1'$ cases, the mass diffusivities are set equal to the thermal diffusivities and the Soret and Dufour cross-diffusion effects are neglected. Additionally these diffusivities are assumed to be constant. Standard forms of Fourier and Fick's law of diffusion are used.
\begin{equation} \label{standard fickian}
   J_j ^{\beta}  = \rho \gamma \frac{\partial Y^{\beta}}{\partial x_j},
\end{equation}
where $\gamma$ represents the Fickian diffusion coefficient. For $'Le \neq 1'$ cases, the diffusivities are calculated from the aforementioned models.
\subsection{Problem geometry and numerical approach}
\begin{figure}
\includegraphics[width=1\textwidth]{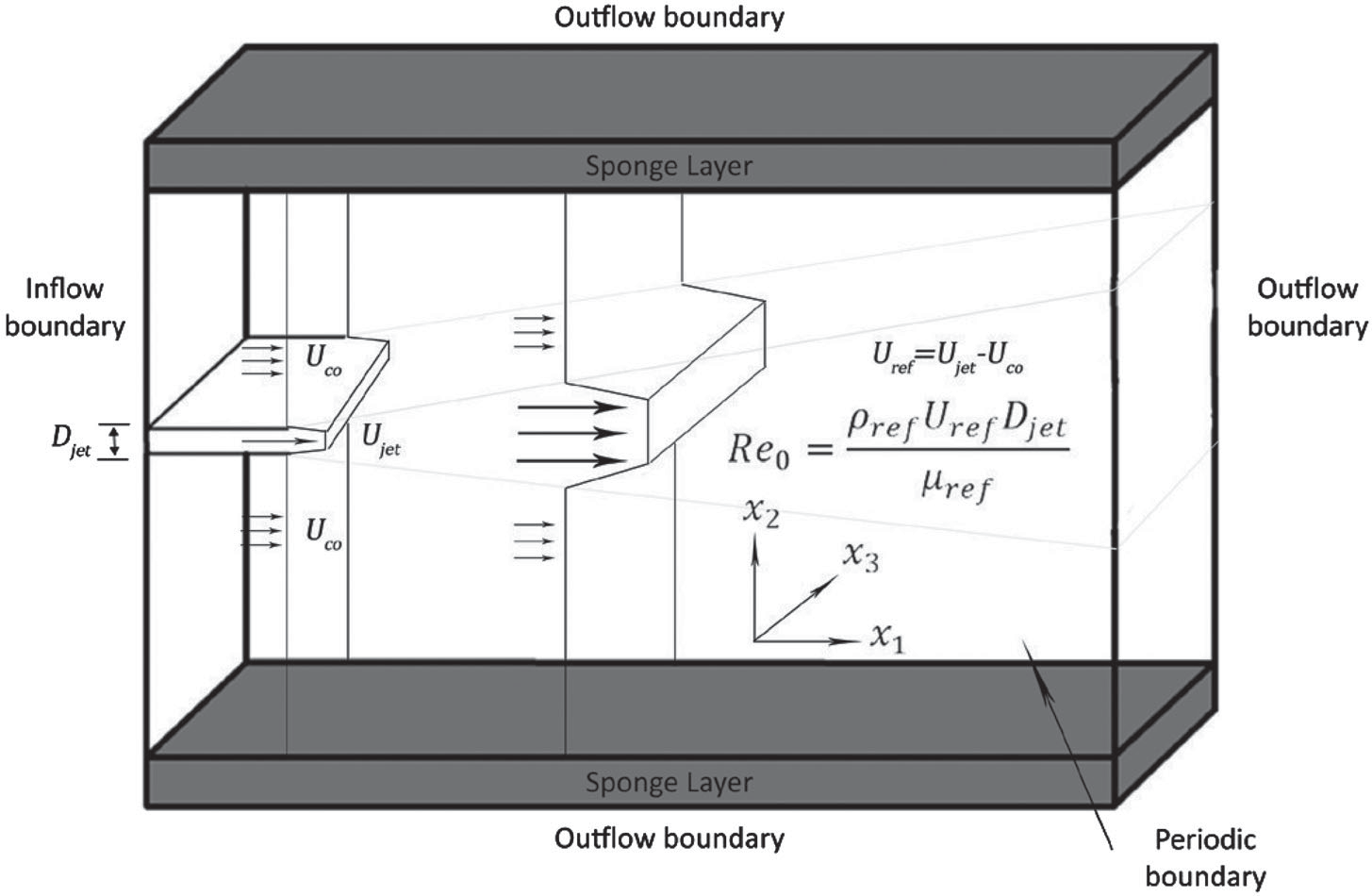}
\caption{Computational domain schematic.}
\label{comput_domain}
\end{figure}

A Cartesian coordinate system is used with $x_1, x_2, x_3$  describing the streamwise, cross-stream and spanwise directions, respectively. The overall domain length in each coordinate direction is chosen such that the flow is fully developed and the modifications to boundary conditions (sponge layer) have no effect on the flow where the statistics are gathered. The domain lengths in the streamwise and cross-stream directions span from $0 < L_1 < 30D_{jet}$ and $-10 D_{jet}< L_2 < 10D_{jet}$, for $Re_0=850$ (nozzle jet width $D_{jet}$, based Reynolds number) and from $0 < L_1 < 45D_{jet}$ and $-15 D_{jet}< L_2 < 15D_{jet}$, for a $Re_0=1300$. The length of the domain in the spanwise direction is adjusted based on the number of grid points in the $x_1$ and $x_3$ directions ($L_3 = (N_3/1.5 N_1) L_1$). This is done to ensure the same grid resolutions in the cross-stream and spanwise directions. A schematic of the computational domain used in the DNS is presented in Fig. \ref{comput_domain}.
%-----------------------
%-----------------------
The governing equations are solved on equally spaced grids in all coordinate directions. Eighth order explicit central finite difference schemes of Kennedy \cite{kennedy} are used for spatial discretization. This central finite difference scheme is closed at the boundary nodes using a scheme called 3-3-4-6-8 \cite{wangetal}. This scheme uses an internally biased third order stencil for nodes 1,2 and fourth and sixth order stencils at boundary nodes 3,4. The boundary stencils are expressed as,
% 33468
\begin{equation} \label{33468}
\begin{split}
f' _1= \frac{1}{6 \Delta x_i} (-11 f_1 + 18 f_2 - 9 f_3 + 2 f_4), \\    
f' _2= \frac{1}{6 \Delta x_i} (-2 f_1 - 3 f_2 + 6 f_3 - f_4), \\
f' _3= \frac{2}{3 \Delta x_i} (f_4 - f_2) - \frac{1}{12 \Delta x_i} (f_5 - f_1), \\
f' _4= \frac{3}{4 \Delta x_i} (f_5 - f_3) - \frac{3}{20 \Delta x_i} (f_6 - f_2) + \frac{1}{60 \Delta x_i} (f_7 - f_1).
\end{split}
\end{equation}
This scheme provides an overall fourth order spatial accuracy near the boundaries and is proven to be stable \cite{wangetal}. The second order derivatives in the viscous and diffusion terms are obtained by applying the first order derivative twice. The temporal derivatives for all the simulations are approximated by a fourth order Runge-Kutta scheme. Tenth order filtering for the spatial derivatives are applied at every Runge-Kutta stage to remove any spurious oscillations in the solutions \cite{kennedy}. CFL conditions for velocities, thermal and mass diffusivities are used to determine the smallest time step.

\begin{figure}
\centering
\subfloat[]{%
\resizebox*{12.5cm}{!}{\includegraphics{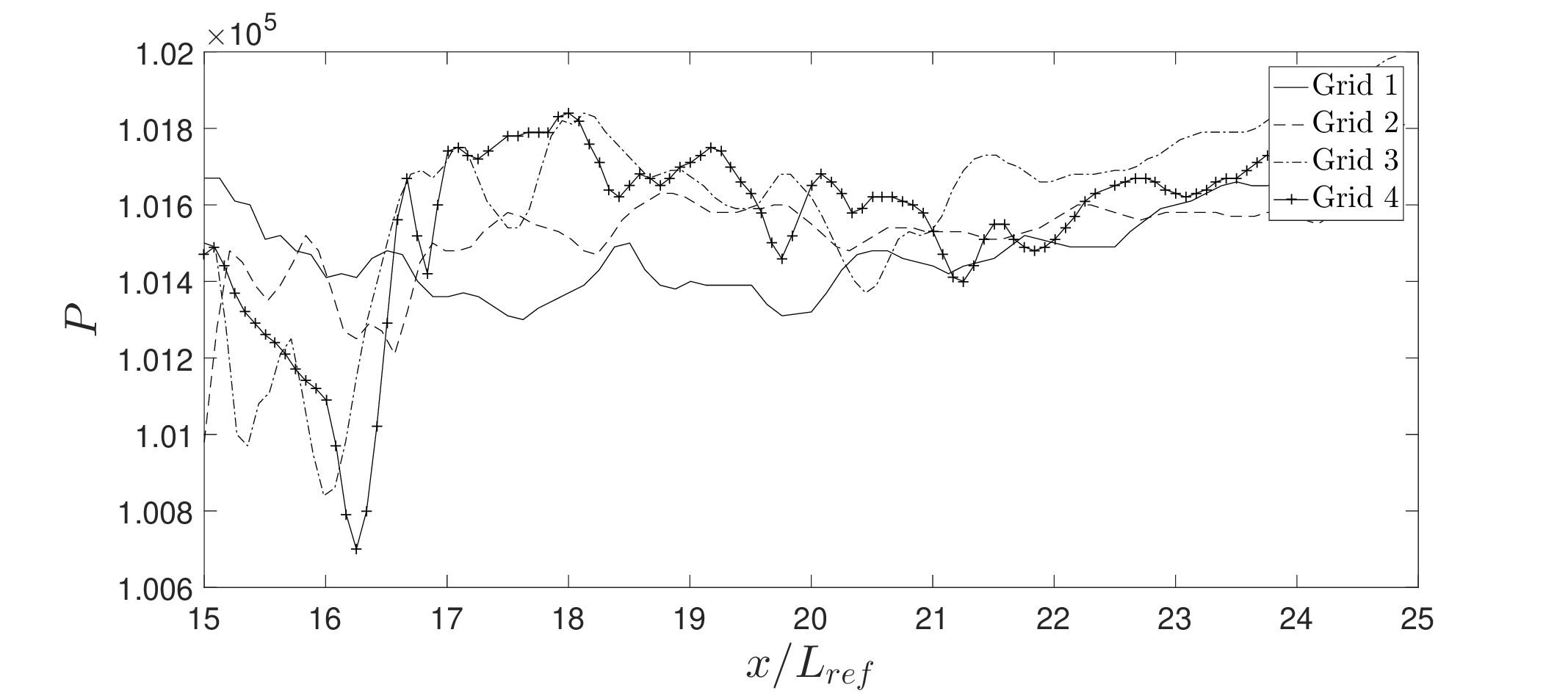}}}\hspace{5pt}
\subfloat[]{%
\resizebox*{12.5cm}{!}{\includegraphics{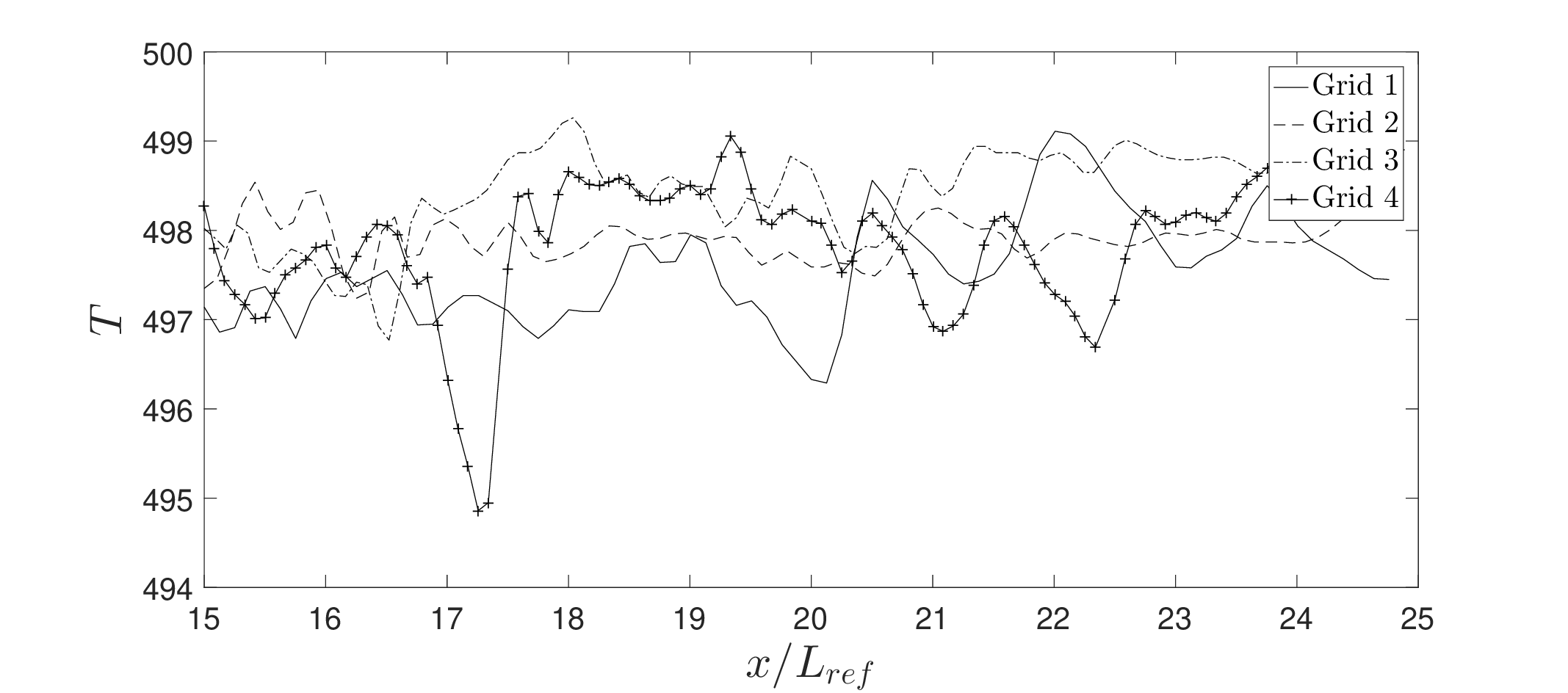}}}
\caption{Properties monitored at various mesh sizes: a) Pressure along jet center line at $Re=850$, b) Temperature along jet center line at $Re=850$.}\label{monprop1}
\end{figure}

\begin{figure}
\centering
\subfloat[]{%
\resizebox*{12.5cm}{!}{\includegraphics{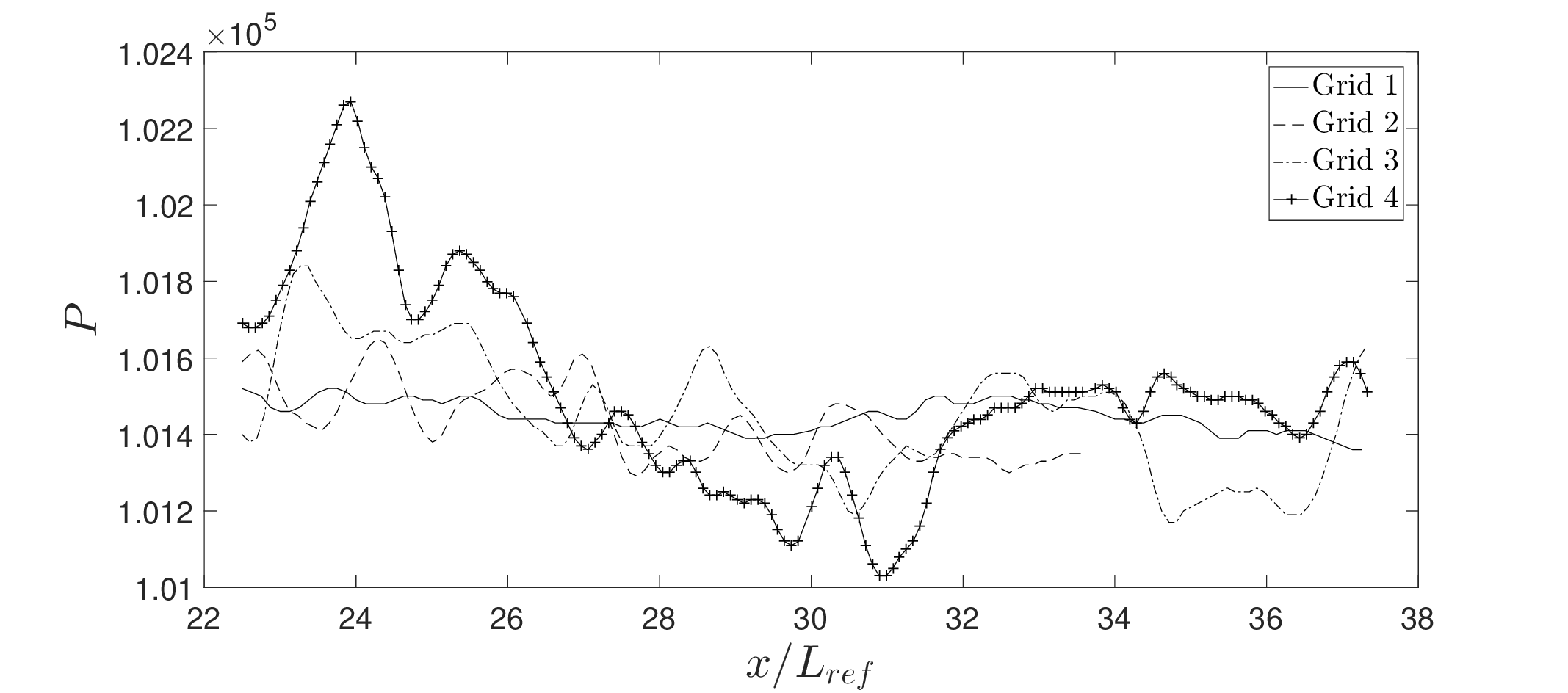}}}\hspace{5pt}
\subfloat[]{%
\resizebox*{12.5cm}{!}{\includegraphics{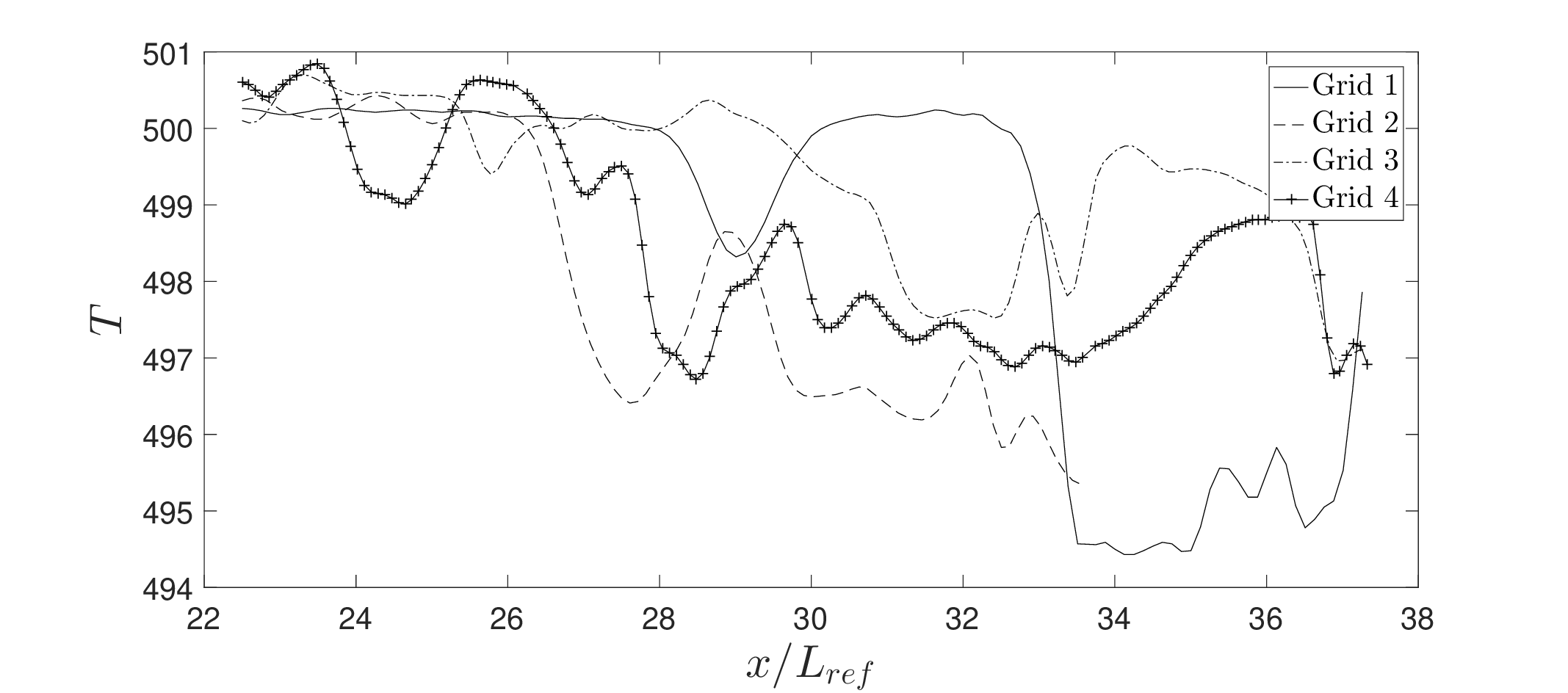}}}
\caption{Properties monitored at various mesh sizes: a) Pressure along jet co-flow at $Re=1300$, b) Temperature along jet co-flow at $Re=1300$.}\label{monprop2}
\end{figure}

Grid resolutions in every coordinate direction are set such that the statistics of the flow become independent of the resolution. A mesh independence study is performed to ensure a valid solution. For each Reynolds number, the simulation is run at an initial mesh size estimated to resolve all length scales. The simulations are run until the convergence criteria is met and the flow properties are monitored at various locations within the domain. The mesh size is then globally refined and the process is repeated. A set of 4 different mesh sizes with increasing resolutions are chosen and tested. With refinement of the mesh, the statistics of the monitored properties are observed to be invariant and the variations in the values of the properties at specific locations are observed to be within the tolerance limit. The smallest of the mesh sizes is then chosen to perform the DNS to reduce the simulation run time. The details of the mesh independence study is presented in the Table \ref{meshindep}. Temperature and pressure measured along the jet center line and co-flow at $Re=850,1300$ are presented in Figs. \ref{monprop1}-\ref{monprop2}.
%---------------- Mesh Independence------------------------
\begin{table}
\tbl{Summary of simulations performed for the mesh independence study. The flow properties are measured along the jet center line and in the co-flow region. The averages of the properties at different mesh sizes are presented. The variations in the values are within the tolerance. All the simulations are performed at $P_0=1 atm$ and convective Mach number of $Ma_c=0.35$.}
{\begin{tabular}{ccccccccccc} \toprule
Grid   & $Re$ & NX  & NY  & NZ  & $\overline{P_{jet}} (Pa)$ & $\overline{P_{co}} (Pa)$ & $\overline{\rho_{jet}}$ & $\overline{\rho_{co}}$ & $\overline{T_{jet}} (K)$ & $\overline{T_{co}} (K)$ \\ \midrule
Grid 1 & 850  & 240 & 240 & 60  & 101480                    & 101540                   & 0.74                    & 0.71                   & 497.6                    & 500                     \\
Grid 2 & 850  & 288 & 288 & 72  & 101520                    & 101570                   & 0.745                   & 0.71                   & 497.9                    & 499.5                   \\
Grid 3 & 850  & 336 & 336 & 84  & 101610                    & 101740                   & 0.74                    & 0.718                  & 498.5                    & 499.9                   \\
Grid 4 & 850  & 360 & 336 & 84  & 101570                    & 101750                   & 0.743                   & 0.72                   & 497.9                    & 499.8                   \\
Grid 1 & 1300 & 360 & 360 & 90  & 101450                    & 101450                   & 0.75                    & 0.72                   & 495.1                    & 498.6                   \\
Grid 2 & 1300 & 432 & 396 & 96  & 101430                    & 101430                   & 0.745                   & 0.72                   & 495.9                    & 498                     \\
Grid 3 & 1300 & 468 & 432 & 100 & 101410                    & 101740                   & 0.74                    & 0.71                   & 499.1                    & 499.4                   \\
Grid 4 & 1300 & 504 & 468 & 120 & 101740                    & 101520                   & 0.74                    & 0.72                   & 501.9                    & 498.5                  \\ \bottomrule
\end{tabular}}
\label{meshindep}
\end{table}

%----------------------------------------------------------
Typical resolutions at the jet centerline are of order $\Delta x_i = 0.083 D_{jet}$ in the $x_2, x_3$ directions and $\Delta x_i = 0.1 D_{jet}$ in the $x_1$ direction. The total number of grid points for this study ranges from 3.456 million to 11.6 million. The code is written in Fortran 77 and parallelized using MPI Fortran subroutines. Table \ref{sims} provides details of the number of grid points and computing resource used for each test case.
%%%%%%%%%%%%%%%%%%%%%%
%---------------- SIMULATIONS ------------------
\begin{table}
\tbl{Summary of simulations at various pressures, diffusion models, Lewis numbers, Reynolds numbers, density ratios, equations of state, simulation run times, grid resolutions and processing cores used. All the simulations are considered at an initial temperature of $T_0 = 500k$ and convective Mach number of $Ma_c=0.35$.}
{\begin{tabular}{lccccccccc} \toprule
Run \# & $P_0$ \textit{(atm)} & Diffusion   & Le       & $Re_0$ & \multicolumn{1}{l}{$\rho_{co}/\rho_{jet}$} & EOS   & $t^*$ & $N_1 \times N_2 \times N_3$ & Cores \\ \midrule
1      & 1                      & Generalized & $\neq 1$ & 850    & 0.558                                     & Real  & 200   & $240 \times 240 \times 60$  & 576   \\
2      & 35                     & Generalized & $\neq 1$ & 850    & 0.519                                     & Real  & 200   & $240 \times 240 \times 60$  & 576   \\
3      & 100                    & Generalized & $\neq 1$ & 850    & 0.479                                     & Real  & 200   & $240 \times 240 \times 60$  & 576   \\
4      & 1                      & Generalized & $\neq 1$ & 1300   & 0.558                                     & Real  & 220   & $360 \times 360 \times 90$  & 600   \\
5      & 35                     & Generalized & $\neq 1$ & 1300   & 0.519                                     & Real  & 220   & $360 \times 360 \times 90$  & 864   \\
6      & 100                    & Generalized & $\neq 1$ & 1300   & 0.478                                     & Real  & 220   & $360 \times 360 \times 90$  & 864   \\
7     & 1                      & Fickian     & $= 1$    & 850    & 0.558                                     & Real  & 200   & $240 \times 240 \times 60$  & 400   \\
8     & 35                     & Fickian     & $= 1$    & 850    & 0.52                                     & Real  & 200   & $240 \times 240 \times 60$  & 400   \\
9     & 100                    & Fickian     & $= 1$    & 850    & 0.479                                    & Real  & 200   & $240 \times 240 \times 60$  & 400   \\
10     & 1                      & Fickian     & $= 1$    & 850    & 0.559                                     & Ideal & 200   & $240 \times 240 \times 60$  & 256   \\
11     & 35                     & Fickian     & $= 1$    & 850    & 0.559                                     & Ideal & 200   & $240 \times 240 \times 60$  & 256   \\
12     & 100                    & Fickian     & $= 1$    & 850    & 0.559                                     & Ideal & 200   & $240 \times 240 \times 60$  & 256   \\
13     & 10                     & Generalized & $\neq 1$ & 850    & 0.564                                     & Real  & 240   & $240 \times 240 \times 60$  & 576   \\
14     & 70                     & Generalized & $\neq 1$ & 850    & 0.493                                     & Real  & 240   & $240 \times 240 \times 60$  & 576   \\ \bottomrule
\end{tabular}}
\label{sims}
\end{table}
%-------------------------------------
%----------mesh independence---------------
%-----
%---------------------------------------

%%%%%%%%%%%%%%%%%%%%%%
\subsection{Initial and Boundary conditions}
The free stream densities $(\rho_{jet}, \rho_{co})$ and speeds of sound $(c_{jet}, c_{co})$ are calculated from the chosen equation of state. The convective Mach number represented by $Ma_c=(U_{jet} - U_{co}) / (c_{jet}+c_{co})=0.35$ for all the simulations in this study. The initial flow Reynolds number ($Re_0=\rho_{ref}U_{ref}D_{jet}/\mu_{ref}$) is calculated based on the reference velocity ($U_{ref}=U_{jet}-U_{co}$), reference  density (average mixture density), reference viscosity (average mixture viscosity) and nozzle width of the jet. The velocities of the jet ($U_{jet}=86.8 m/s$) and co-flow ($U_{co}=17.4 m/s$) are chosen such that the transition to turbulence due to shear effects at the interface is maximized for the chosen forcing frequency. $D_{jet}$ is changed to set or change the $Re_0$ in the different cases in the Table \ref{sims}. In order to obtain a stationary flow, the simulation is run for a  non-dimensional time ($t^* = t U_{ref}/D_{jet}$) of 200-240 (Table \ref{sims}) to obtain converged statistics.

This code makes use of the form of NSCBC proposed by O'kong and Bellan \cite{bellanbc} along with a sponge layer in the $x_2$ direction. The formulation of NSCBC includes real gas effects. The sponge layer is similar to that of Wasistho \textit{et al.} \cite{sponge} where a given conservative variable is steadily brought back to its reference value. This reduces the intensity of outgoing waves through the boundary. It may be expressed as,
\begin{equation}
U_{c,i} = U_{c,ref} + f_{1} (U_{c,i}-U_{c,ref})
\end{equation}
\begin{equation}
f_{1} = [1-C_1 f_{2} ^2] \frac{1-(1-e^{C_2 f_{2} ^2})}{1-e^{C_2}}
\end{equation}
\begin{equation}
f_{2} = \frac{x_i - x_{start}}{x_{end}-x_{start}}
\end{equation}
In the above equations, $U_{c,i}$ represents the conservative variables in the governing equations, $U_{c,ref}$ the reference value for the corresponding conservative variable obtained from the free stream, $C_1, C_2$ the adjustable model constants to change the strength of the sponge zone, and $x_{start}, x_{end}$ the start and end of the sponge zone, respectively.

In the $x_1$ direction, a subsonic non-reflecting outflow boundary condition is used where the convection velocities for all characteristic waves other than pressure is increased by a factor of the local speed of sound ($\lambda_i + c$, i=2,...N). The amplitude of incoming waves is fixed at $C_K (p-p_{\infty})$. The coefficient $C_K$ as proposed by Rudy and Strikwerda \cite{rudystrikwerda} sets the intensity of incoming waves and $p_{\infty}$ represents the ambient pressure inside the computational domain. This boundary condition has an effect similar to a supersonic boundary condition where all the flow properties are convected out of the domain without any reflection, but at the same time, it also allows for stabilization of the mean pressure within the domain. Since this is applied only at the face of the outflow boundary, it does not cause stretching of the turbulent structure near boundaries or affect the incoming pressure wave. The flow region upstream of this boundary is observed to have no effects due to presence of this boundary condition. The sponge zones used in the $x_2$ direction prevent interaction of flow with corners of the computational domain at the outflow region and so prevent any reflection from the corners. Periodic boundary condition is used in the $x_3$ direction. A subsonic non-reflecting inflow boundary condition \cite{poinsotlele} is used at the inlet in the $x_1$ direction and is determined to be stable for mixing flows.
\subsection{Inflow condition}
To resolve the flow profile at the inflow, the streamwise velocity in the shear layer on either side of the jet is given by a hyperbolic tangent function. It may be expressed as,
\begin{equation}
u(x_2) = \frac{U_{jet}+U_{co}}{2} - \frac{U_{ref}}{2} tanh \Big(\frac{|x_2|-0.5D_{jet}}{2\theta_0}\Big),
\end{equation}
where $\theta_0=0.05D_{jet},$ is the shear layer momentum thickness, and $U_{ref} = U_{jet}-U_{co}$. This results in a symmetric top-hat velocity profile with smooth edges about the center-line. The cross-stream and spanwise velocities are specified as zero at the inflow. The initial mean pressure and temperature are initialized to be uniform throughout the domain. The mean profile for the species mass fractions is also given by a hyperbolic tangent function, 
\begin{equation}
\begin{split}
Y_{fuel} = 1 - tanh(\frac{|x_2|}{2\theta_0}),\\
Y_{ox} = 1 - Y_{fuel}.
\end{split}
\end{equation}
For multicomponent mixing like in the case of heptane-air, the intial mass fractions of oxygen and nitrogen are specified as $Y_{O_2}=0.233 Y_{ox}, Y_{N_2}=0.767 Y_{ox}$.

For turbulent inflow generation, the forcing method used combines the sinusoidal perturbation method with a circular spatial forcing term to induce spatio-temporal instability. The forcing term is expressed as,
\begin{equation}
\begin{split}
u_1(t)=u_1(t)+A_f[Sin(2\pi ft) \theta], \\
\theta = tan^{-1}(\frac{x_3}{x_2}),
\end{split}
\end{equation}
where $A$ represents the amplitude of forcing signal which is chosen to be a value between 0.001-0.002 (larger amplitudes cause production of stronger numerical waves that corrupt the flow, while perturbations with smaller amplitudes get dampened quickly) and $f$ represents the most unstable frequency which determines the distance away from the inlet at which flow transition takes place. Based on a number of simulations performed, a value of $f$ between 0.25-0.4 is chosen such that the distance of flow transition is less than $x_1=10 D_{jet}$ from inlet. However, these perturbation frequencies work only for velocities ($U_{jet} = 86.8 m/s, U_{co} = 17.4 m/s$). For flows with different velocities, the most unstable frequency may be determined from linear stability analysis \cite{lsaspatial}. The circular spatial forcing term $\theta=\theta(x_2,x_3)$ induces three dimensional spatial instability. Instability in spatial flows is induced when the wave number is complex and wave frequency is real ($u'(y,t)=u(y) \exp{i(kx-\omega t)}$, with $k$ - complex, $\omega$ - real). Since the boundaries in the spanwise directions are periodic, a forcing method without $\theta$ does not induce cascading. The flow evolution and the flow statistics measured from the DNS are presented in Section \ref{dnsresults}
\section{FMDF method}
The filtered mass density function (FMDF) method is a stochastic approach to modelling turbulent flows in compressible LES. The FMDF method was proposed by Jaberi \textit{et al.} \cite{fdfintro,fmdfintro} and is based on the PDF method for reactive flows proposed by Pope \cite{popereac}. The idea of this method is to obtain a complete statistical description of the flow from a joint PDF defined for the scalars and turbulent flow variables. A transport equation for this joint PDF is then derived, from which the statistical moments of the flow can be determined. The main advantage of this method is that the reaction source term in the energy and species equations remains closed and requires no modelling. The FMDF method is similar to the PDF method except that the FMDF performs a Favre filtering operation on the PDF. For a scalar $\phi_{\alpha}(x,t)$, an exact transport equation for the scalar-FMDF ($F_L$) can be derived following the procedures of Pope \cite{popereac}, Colucci \textit{et al.} \cite{fdfintro}, Jaberi \textit{et al.} \cite{fmdfintro}. The final form of the FMDF transport equation may be expressed as
\begin{equation} \label{fmdftransport}
\frac{\partial F_L}{\partial t} + \frac{\partial \lbrack \langle u_i | \psi \rangle_l F_L \rbrack}{\partial x_i} = \frac{\partial}{\partial \psi_{\alpha}} \Big \lbrack \Big \langle \frac{1}{{\rho}} \frac{\partial J_i ^{\alpha}}{\partial x_i} \Big| \psi \Big \rangle_l F_L \Big \rbrack - \frac{\partial \lbrack S_{\alpha} F_L \rbrack}{\partial \psi_{\alpha}}
\end{equation} % working
The chemical reaction source term in the above equation is closed. However, two new terms appear that remain unclosed: $\langle u_i | \psi \rangle_l$, in the second term on the left hand side of the equation is called subgrid scalar convection and represents the motion of scalars at the subgrid scales due to the flow velocity. This term may be decomposed as
\begin{equation}
\langle u_i | \psi \rangle_l F_L = \langle u_i \rangle_L F_L + \lbrack \langle u_i | \psi \rangle_l - \langle u_i \rangle_L \rbrack F_L,
\end{equation}
where the term $\lbrack \langle u_i | \psi \rangle_l - \langle u_i \rangle_L \rbrack F_L$ represents the SGS convective flux and may be modeled using a functional model: 
\begin{equation} \label{sgsconv}
\lbrack \langle u_i | \psi \rangle_l - \langle u_i \rangle_L \rbrack F_L = -\gamma_t \frac{\partial (F_L / \langle \rho \rangle_l)}{\partial x_i}.
\end{equation}
Equation (\ref{fmdftransport}) can then be rewritten as, 
\begin{equation} \label{fmdftrans2}
\frac{\partial F_L}{\partial t} + \frac{\partial \lbrack \langle u_i \rangle_L F_L \rbrack}{\partial x_i} = \frac{\partial}{\partial x_i} \Big \lbrack (\gamma+\gamma_t) \frac{\partial (F_L / \langle \rho \rangle_l)}{\partial x_i} \Big \rbrack + \frac{\partial}{\partial \psi_{\alpha}} \Big \lbrack \Big \langle \frac{1}{{\rho}} \frac{\partial J_i ^{\alpha}}{\partial x_i} \Big| \psi \Big \rangle_l - S_{\alpha} \Big \rbrack F_L.
\end{equation}
This transport equation is solved using a Lagrangian Monte Carlo method where the spatial transport is governed by a stochastic differential equation and the composition transport is solved in a mesh free Lagrangian setup simultaneously. The composition evolution of a Lagrangian fluid element ($\phi_{\alpha}$) can be written as,
\begin{equation}
    \frac{d\phi_{\alpha}}{dt} = \Theta_{\alpha} - S_{\alpha},
\end{equation}
with $\Theta_{\alpha} = \Big \langle \frac{1}{{\rho}} \frac{\partial J_i ^{\alpha}}{\partial x_i} \Big| \psi \Big \rangle_l$. This term represents the conditional scalar diffusion (CSD). It remains as the last unknown term in this transport equation but treatment of this term using a functional approach is not straightforward. It represents the effects of molecular transport (diffusion) in physical space and molecular mixing in composition space. A few mixing models that stem from different physical concepts have been proposed for treatment of this term. The following section briefly discusses the Interaction by Exchange of Mean, Modified Curl, and Mapping Closure methods for modelling the CSD.
\subsection{Molecular mixing models}
\subsubsection{Interaction by Exchange of Mean} \label{iem_intro}
The Interaction by Exchange of Mean (IEM) \cite{IEMvillermaux}, otherwise known as the Linear Mean Square Estimation (LMSE) \cite{dopazobrien}, is one of the simplest deterministic models that describes the phenomena of mixing at the subgrid scale. The composition evolution of subgrid scale particles in this model is given by,
\begin{equation}
    \frac{d\phi_{\alpha}}{dt} = \Theta_{\alpha} = \Omega_m (\psi_i - \langle \psi \rangle_l),
\end{equation}
where $\langle \psi \rangle_l$ represents the volume average of a scalar $\psi$ and $\psi_i$ represents the instantaneous value of the scalar within the filter confines. In essence, the model causes the composition of a Lagrangian particle within a given filter volume to relax towards the mean value at a rate $\Omega_m \sim \varepsilon / k$, where $\Omega_m$ represents the mixing frequency, $\epsilon$ the dissipation rate and $k$ the turbulent kinetic energy. The IEM model is local only in physical space but the conditional scalar diffusion it models is local in both spatial and composition space. Thus, for dispersed flows that have large spatial variation of concentration, the IEM model may not be well suited. 
\subsubsection{Interaction by Exchange of Conditional Mean}
An improvement to the IEM model, called Interaction by Exchange of Conditional Mean (IECM) was proposed by Pope \cite{IECM1} and Fox \cite{IECM2} to overcome the dispersion inconsistency \cite{spmmpope}. In this model, instead of standard averaging of the scalar (volume average within filter), a velocity conditioned average of the scalar is used. In the IEM model, the particles that have similar spatial positions interact with each other but in the IECM model the particles that have similar velocities interact with each other. Physically it represents interactions of particles that belong to the same eddy \cite{bakosi}. Thus, in dispersed turbulent flows, the IECM model is expected to be more accurate than IEM in predicting the SGS mixing phenomena. The IECM model is expressed as,
\begin{equation}
\Theta_{\alpha} = \Omega_m (\phi_i - \langle \phi | U \rangle_l).
\end{equation}
However, this model is also non-local in composition space.
\subsubsection{Modified Curl} \label{modifiedcurl}
The modified curl (MC) model proposed by Janicka \textit{et al.} \cite{janickaetal} is a particle interaction model based on Curl's \cite{curlmc} model. For equally weighted particles, pairs of particles are randomly selected from an ensemble and their compositions are changed as,
\begin{equation}
\begin{split}
    \phi_{i,new} = \phi_i + 0.5 R_{ij} (\phi_j - \phi_i), \\
    \phi_{j,new} = \phi_j + 0.5 R_{ij} (\phi_i - \phi_j).
\end{split}
\end{equation}
The model for conditional diffusion can be re-written as,
\begin{equation} \label{modcurl}
\begin{split}
\Theta_{\alpha} = \Omega_m R_{ij} C_{MC} (\phi_c - \phi_j), \\
\phi_c = \frac{w_i \phi_i + w_j \phi_j}{w_i + w_j},
\end{split}
\end{equation}
where $\phi_i,\phi_j$ represent particles from the ensemble, $w_i, w_j$ represents the weights of the particles, $R_{ij}$ represents a uniformly distributed random number, $C_{MC}$ represents the model constant and $\Omega_m$ is mixing frequency similar to one used in the IEM model. A model for unequal weights was developed by Nooren \textit{et al.} \cite{noorenetal} but effective working of model depends on distribution of particle weights. Unlike the IEM, the MC model is stochastic in nature but is still local in physical space and not in composition space.
\subsubsection{Mapping Closure} \label{mapping}
The Mapping Closure (MAPPING) model for CSD was formulated by Kraichnan \cite{map1} and Chen \textit{et al.} \cite{map2}. The model suggests that the scalars in turbulent flows follow an assumed distribution. A model for mixing of particles that follow this distribution is then derived. For unequally weighted particles, the model proposed by Pope \cite{mappope} maps the composition space of a single scalar with a Gaussian reference field. The scalar evolution of particles can be written as, 
\begin{equation}
\begin{split}
    \frac{d\phi_1}{dt} = - C_{map} \Omega_m [B_{1+\frac{1}{2}} (\phi_2 - \phi_1)], \\
    \frac{d\phi_i}{dt} = C_{map} \Omega_m [B_{i+\frac{1}{2}} (\phi_{i+1} - \phi_i)-B_{i-\frac{1}{2}} (\phi_{i} - \phi_{i-1})]; i=2,...N_p-1,  \\
    \frac{d\phi_{N_p}}{dt} = - C_{map} \Omega_m [B_{N_{p}-\frac{1}{2}} (\phi_{N_p} - \phi_{N_{p}-1})],
\end{split}
\end{equation}
where $C_{map}$ represents the model constant and $\Omega_m$ the mixing frequency. The scalars $\phi_i$ in the equations are sorted from the smallest to largest values ($\phi_i < \phi_{i+1}$). The coefficients $B$'s are defined as,
\begin{equation} \label{eq_b}
\begin{split}
    B_{i+\frac{1}{2}} = \frac{N_p g(\eta_{i+\frac{1}{2}})}{\eta_{i+1}-\eta_i}, \\
    B_{i-\frac{1}{2}} = \frac{N_p g(\eta_{i-\frac{1}{2}})}{\eta_{i}-\eta_{i-1}}.
\end{split}
\end{equation}
$N_p$ in this equation represents the total number of particles and $g(\eta)$ represents a standard Gaussian PDF with sample space coordinate $\eta$ defined as,
\begin{equation}
\begin{split}
   \eta_{i+\frac{1}{2}} = G^{-1} (p_i), \\
   \eta_i = G^{-1} (p_{i-\frac{1}{2}}), \\
   \delta \eta_i  = \eta_{i+\frac{1}{2}} - \eta_i, \\
\end{split}
\end{equation}
where $\delta \eta_i$ is the half step difference such that $\eta_{i+1} = \eta_{i+\frac{1}{2}} + \delta \eta_i$ and $G^{-1}$ represents the inverse cumulative density function corresponding to the PDF $g$. The arguments $p$'s are defined as,
\begin{equation}
\begin{split}
    p_{i+\frac{1}{2}} = \frac{\sum_{j=1} ^i w_j - 0.5 w_i}{W}, \\
    p_i = \sum_{j=1} ^i w_j / W,
\end{split}
\end{equation}
where $W$ in the equation above represents the sum of weights of all particles in the ensemble. The boundary values of coefficients are given as,
\begin{equation}
\begin{split}
    B_{1+\frac{1}{2}} = \frac{N_p g(\eta_{1+\frac{1}{2}})}{\eta_2-\eta_1},\\
    B_{N_p-\frac{1}{2}} = \frac{N_p g(\eta_{N_{p}-\frac{1}{2}})}{\eta_{N_p}-\eta_{N_p -1}},\\
    \eta_{1+\frac{1}{2}} = \eta_1 + \delta \eta_1, \\    
    \eta_{N-\frac{1}{2}} = \eta_{i+\frac{1}{2}} |_{i=N_p-1}.
\end{split}
\end{equation}
The mapping closure model preserves the mean and variance decay. Since the scalars are sorted, the resulting model is local in composition space. However, the model is applicable only to a single scalar. An extension of this model to multiple scalars called the Euclidean Minimum Spanning Tree (EMST) was proposed by Pope \cite{emst}. However, it is not studied in this paper.
\subsubsection{Mixing frequency}
The term $\Omega_m$ in the aforementioned models represents the mixing frequency. The mixing frequency can be determined from the parameters measured in the simulation using a dimensional analysis. Jaberi \textit{et al.} \cite{fmdfintro} proposed an expression for $\Omega_m (x,t)$ as,
\begin{equation} \label{mixingfreq}
    \Omega_m = C_{\Omega} \frac{(\gamma + \gamma_t)}{(\langle \rho \rangle_l \Delta_G ^2)}.
\end{equation}
In this equation, $\gamma$ represents the molecular diffusivity, $\gamma_t=\langle \rho \rangle_l \frac{C_R \Delta_G \sqrt{\xi}}{Sc_t}$ the subgrid scalar diffusivity, $\langle \rho \rangle_l$ the filtered density, $\Delta_G$ the filter width, $Sc_t$ the subgrid Schmidt number, and $C_R, C_{\Omega}$ the model constants. The standard model constants for atmospheric pressure flows are obtained from the works of Jaberi \textit{et al.} \cite{fmdfintro}. The term $\xi$ is expressed as,
\begin{equation}
\begin{split}
\xi = |\langle u_i ^* \rangle_L \langle u_i ^* \rangle_L - \langle \langle u_i ^* \rangle_L \rangle_{l'} \langle \langle u_i ^* \rangle_L \rangle_{l'}|, \\
    u_i ^* = u_i - U_{ref,i}.
\end{split}
\end{equation}
The dimension of the mixing frequency is $[1/s]$.
\subsection{Measurement of subgrid scale terms from DNS}
The subgrid flow phenomena that is modeled in LES can be calculated from DNS by breaking down the DNS domain and performing volume averages of flow properties within each of the smaller domains.
\begin{equation}
\begin{split}
    L_i=N_{i,D} \Delta x_{i,D} = N'_{i,L} \Delta x'_{i,L}, \\
    \Delta x'_{i,L} >> \Delta x_{i,D}, \\
    N_{i,D} >> N'_{i,L},
\end{split}
\end{equation}
where $N_{i,D}, N'_{i,L}$ represent the number of grid points, and $\Delta x_{i,D}, \Delta x'_{i,L}$ represent the grid size in a coordinate direction used in DNS and LES, respectively. The subgrid flow characteristics (filtered mass density function, conditional scalar diffusion) are directly computed from the DNS and compared against the SGS models to evaluate the effectiveness of closure at large pressures with a generalized diffusion model.
\subsubsection{Measurement of FMDF}
The FMDF represents the distribution of fluctuations of subgrid scalars. Unlike a PDF in RANS, the FMDF is stochastic in nature. The PDF in RANS characterizes fluctuations over various flow realizations, but the FMDF characterizes the instantaneous subgrid fluctuations. The mean of the LES-FMDF is equal to the RANS-PDF when the filter width tends to zero \cite{wangfdf}. Inspite of the differences between the two, the FMDF has all the mathematical properties of a PDF. Thus, in LES the FMDF may be used to obtain all the statistics of the flow. The FMDF for a scalar can be obtained from the DNS data by applying a spatial filter of appropriate width to the scalar variable in the DNS code, extracting the filtered scalar at various points in the flow and obtaining the instantaneous PDF of the filtered scalar.
\subsubsection{Measurement of CSD}
In order to simplify the analysis, only statistics of the FMDF of the mixture fraction are studied. For the DNS with generalized diffusion models (without an effective diffusion coefficient) a transport equation for mixture fraction is required to obtain the scalar diffusion rate. The mixture fraction definition used in this work may be expressed as, 
\begin{equation} \label{mixfrac}
    \phi = \frac{Y_f - Y_o + Y_o ^{0}}{Y_f ^{0} + Y_o ^{0}},
\end{equation}
where $Y_f, Y_o$ represent the instantaneous mass fractions and $Y_f ^{0}, Y_o ^{0}$ represent the initial free stream mass fractions of fuel and oxidizer, respectively. For heptane-air mixture, the mass fraction of oxidizer indicates the sum of mass fractions of oxygen and nitrogen. The mixture fraction may take a value between 0 (pure oxidizer) and 1 (pure fuel) indicating the proportion of fuel within the mixture. For multi-species mixtures, the mixture fraction should ideally be defined in terms of an elemental mass fraction \cite{justinPhD}. However, for pure mixing cases and for the purposes of this study, the definition in Eq.(\ref{mixfrac}) is sufficient. The mixture fraction expression may be rewritten as,
\begin{equation} \label{mfrac2}
    Y_f - Y_o = \phi (Y_f ^{0} + Y_o ^{0}) - Y_o ^{0}.
\end{equation}
The transport equations for fuel and oxidizer [Eq.(\ref{species})] may be expressed as,
\begin{equation} \label{ftr}
        \frac{\partial}{\partial t} (\rho Y_f) + \frac{\partial}{\partial x_j} [\rho u_j Y_f + J_{j,f}] = 0,
\end{equation} 
\begin{equation} \label{otr}        
        \frac{\partial}{\partial t} (\rho Y_o) + \frac{\partial}{\partial x_j} [\rho u_j Y_o + J_{j,o}] = 0.
\end{equation}
Subtracting Eqs.(\ref{otr}) from (\ref{ftr}) gives,
\begin{equation} \label{diffj}
       \frac{\partial \rho}{\partial t} (Y_f - Y_o) + \frac{\partial}{\partial x_j} [\rho u_j (Y_f - Y_o) + J_{j,f} - J_{j,o}] = 0.
\end{equation}
Substituting the mixture fraction definition from Eq.(\ref{mfrac2}) into Eq.(\ref{diffj}) gives,
\begin{equation}
\begin{split}
      \frac{\partial \rho}{\partial t} [\phi (Y_f ^{0} + Y_o ^{0}) - Y_o ^{0}] + \frac{\partial}{\partial x_j} \Big \lbrack \rho u_j [\phi (Y_f ^{0} + Y_o ^{0}) - Y_o ^{0}] + J_{j,f} - J_{j,o} \Big \rbrack = 0, \\
      (Y_f ^{0} + Y_o ^{0}) \frac{\partial \rho \phi}{\partial t} - Y_o ^{0}  \cancelto{0}{\Big \lbrack \frac{\partial \rho}{\partial t} + \frac{\partial \rho u_j}{\partial x_j} \Big \rbrack} + (Y_f ^{0} + Y_o ^{0}) \frac{\partial}{\partial x_j} (\rho u_j \phi) + \frac{\partial}{\partial x_j} (J_{j,f} - J_{j,o}) = 0.
\end{split}      
\end{equation}
The second term in the above equation results in the continuity equation. The final transport equation for mixture fraction can then be written as,
\begin{equation}
\frac{\partial \rho \phi}{\partial t} + \frac{\partial}{\partial x_j} (\rho u_j \phi) + \frac{1}{(Y_f ^{0} + Y_o ^{0})} \frac{\partial}{\partial x_j} (J_{j,f} - J_{j,o}) = 0,
\end{equation}
where the scalar diffusion rate for mixture fraction with generalized diffusion model is expressed as,
\begin{equation}
    \Big( \frac{\partial J_j ^{\phi}}{\partial x_j} \Big)_{G} = \Big \lbrack \frac{1}{(Y_f ^{0} + Y_o ^{0})} \frac{\partial}{\partial x_j} (J_{j,f} - J_{j,o}) \Big \rbrack
\end{equation}
To measure the CSD, the scalar diffusion rate of mixture fraction is calculated throughout the domain in the DNS code. At a spatial location within the flow, an appropriate filter size is chosen ($\Delta_f = r \Delta x_i$, where $r$ is the radius of a sphere for a spherical filter or half the length of a side for a cubic filter) for measurement of subgrid terms. A filter of volume defined by the filter size is then constructed ($V_f = \frac{4}{3} \pi \Delta_f ^3$, for spherical filter and $V_f=\Delta_f ^3$ for cubic filter). The terms (scalar diffusion and mixture fraction) at grid nodes within the filter confines are then extracted. A discrete set of grid points that fall within the filter volume can be thought of as FMDF particles. The averages of scalar diffusion rate of the subgrid particles conditioned at various mixture fractions are then obtained. In this work a cubic filter is used and the ratios of filter width to grid size $(r= \Delta_f / \Delta x_i)$ are chosen as 4 and 16. A similar procedure may be applied to obtain conditional averages of other scalars or scalar diffusion conditioned at different values of other variables (velocity conditioned diffusion, velocity-scalar conditioned diffusion).
%%%%%%%%%%%%%%%%%%%%%%%%%%%%%%%%%%%%%%%%%%%%%%%%%%%%%%%%%%%%%%%%%%%
\section{Results}
\subsection{DNS flow evolution} \label{dnsresults}
\subsubsection{Structural Evolution of Jet}
\begin{figure}
\includegraphics[width=0.9\linewidth]{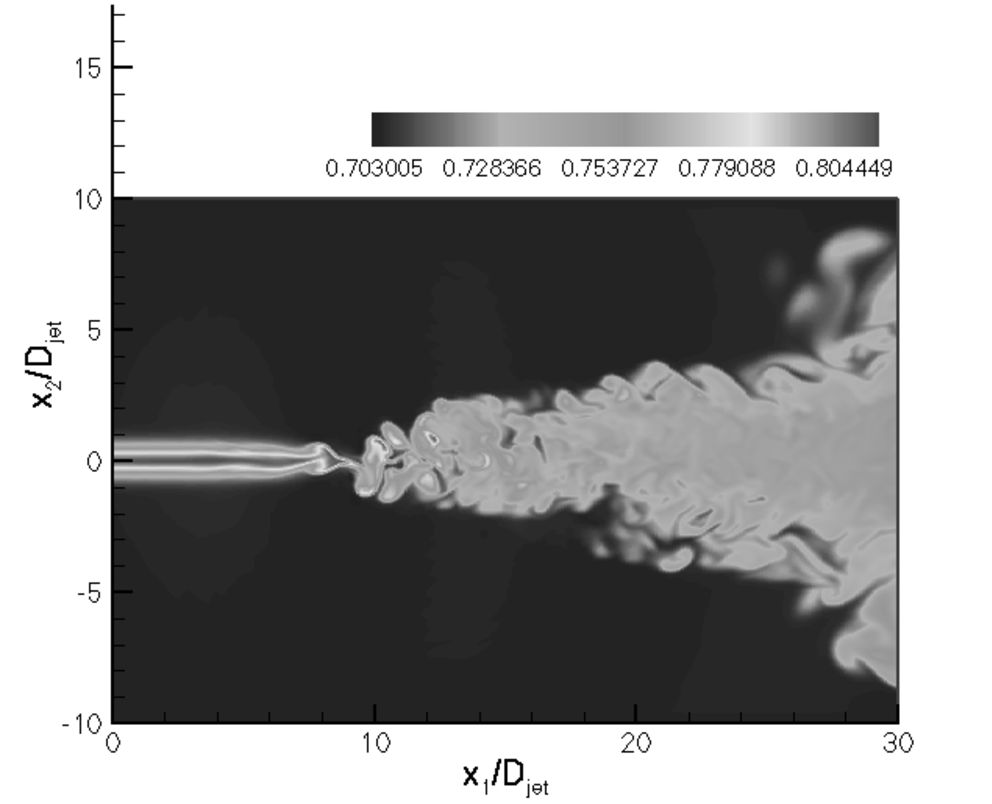}
\caption{Instantaneous density contour on major plane at $P_0=100 atm$, $Re_0=850$.}
\label{contourden}
\end{figure}

\begin{figure}
\includegraphics[width=0.9\linewidth]{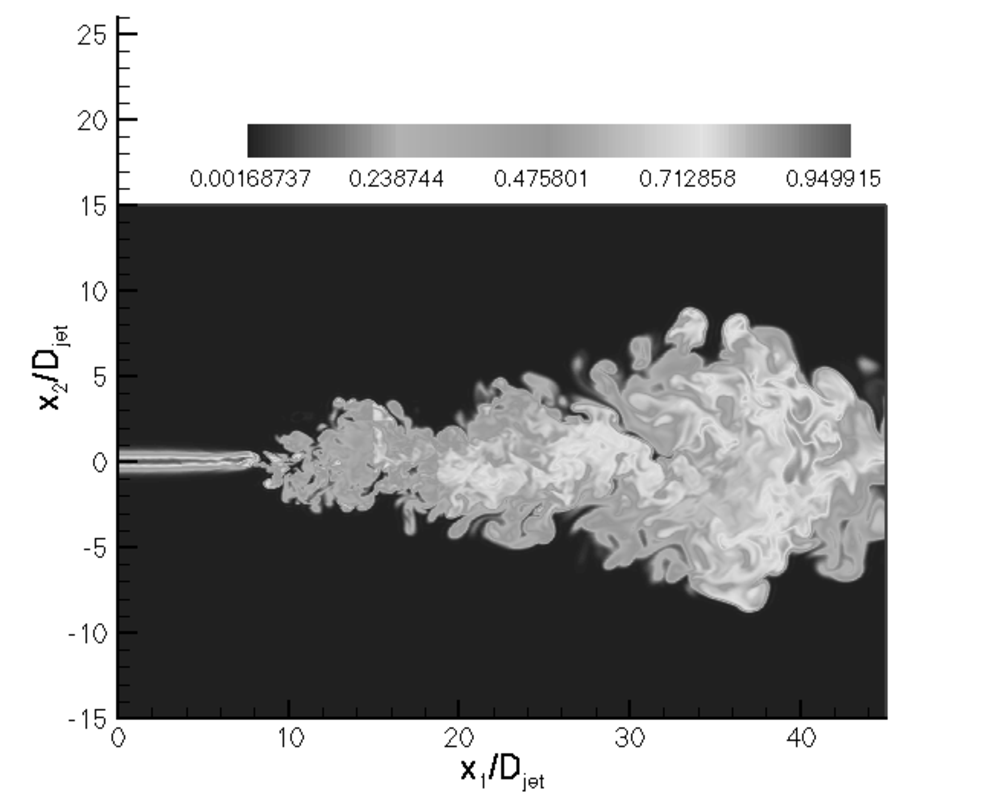}
\caption{Mixture fraction contours in major jet plane at $P_0=100 atm$, $Re_0=1300$.}
\label{contourmixf}
\end{figure}

The simulation is started with a laminar flow profile throughout the domain and the inflow perturbation is then turned on. The transition process may be visualized with instantaneous two dimensional contour plots of density in Fig. \ref{contourden} and mixture fraction in Fig. \ref{contourmixf}. It is observed that in the region $x_1 < 10 D_{jet}$, the flow is laminar and roll up of vortex rings due to Kelvin Helmholtz instability \cite{khins} is observed in the region $10 D_{jet} < x_1 < 15 D_{jet}$. These vortices represent emergence of varicose modes from the sinusoidal perturbations. These vortex rings move downstream and begin to pair. This corresponds to natural subharmonic perturbations generated after formation of primary vortices. Further downstream, these structures breakup and indicate the cascading effect. Around $x_1 > L_x/2$, the jet potential core ends and the flow becomes turbulent.
%------------------
%-----------------
%------------------
%%%%%%%%%%%
\subsubsection{Mean Velocity of Flow}
\begin{figure}
\centering
\subfloat[]{%
\resizebox*{12.5cm}{!}{\includegraphics{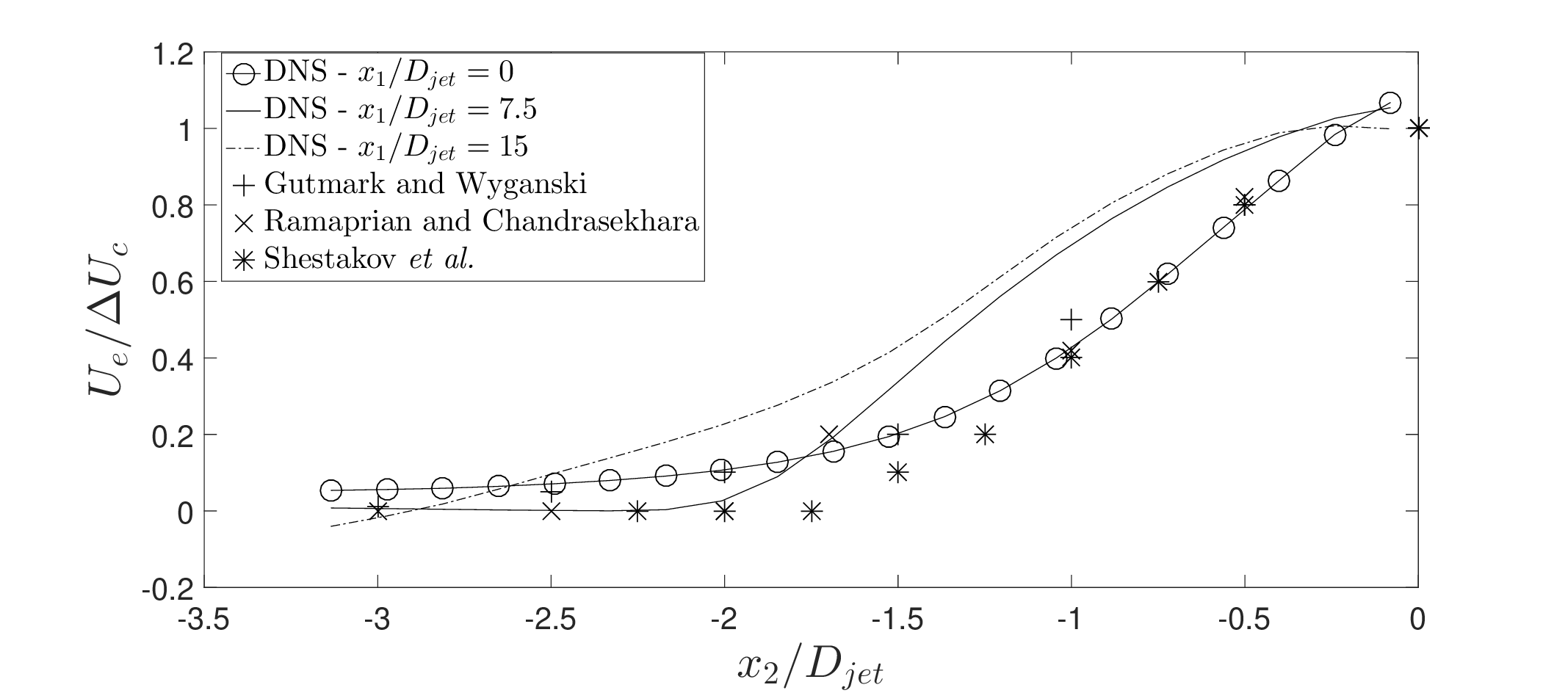}}}\hspace{5pt}
\subfloat[]{%
\resizebox*{12.5cm}{!}{\includegraphics{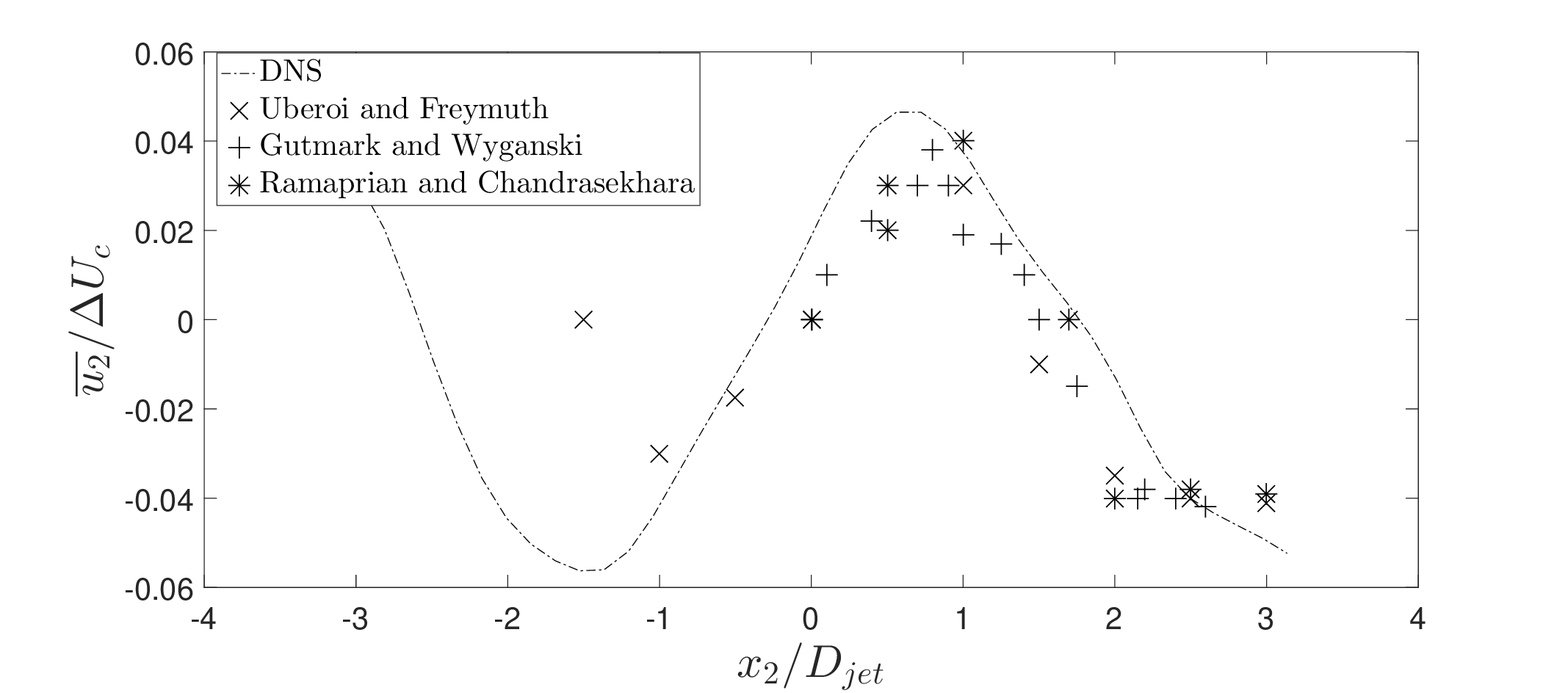}}}
\caption{Mean velocity excess profiles in slot jet: a) Streamwise velocity, b) Cross-stream velocity.}\label{mve}
\end{figure}

Time averaged statistics of flow variables are calculated and compared against experimental results to validate the DNS data. Although the flow profile in this work is a slot jet, the flow evolution is structurally similar to plane jets. Thus, the flow evolution is validated against experimental results from both slot jets and plane jets. Mean velocity excess in jets describe the spatial evolution of the flow. The normalized streamwise mean velocity excess at various downstream locations are presented in Fig. \ref{mve}. The term $U_e = \overline{u_1(x_2)} - \overline{U_{co}}$ denotes the mean streamwise velocity excess, $\overline{u_1(x_2)}$ the local mean velocity, $\overline{U_{co}}$ the local mean co-flow velocity and $\Delta U_c$ the difference between local mean centerline velocity and local mean coflow velocity. The mean streamwise velocity is observed to collapse to a semi self similar profile after a distance $x_1 > 7.5 D_{jet}$, from the inlet. This shows good agreement with data from experiments on slot jets by Shestakov \textit{et al.} \cite{slotjet} and plane jets by Gutmark and Wygnanski \cite{gutwyg}, and Ramaprian and Chandrasekhara \cite{ramchan}. Although the physical models used in this simulation differ from the experiment, the flow parameters and evolution can be compared as long as the Reynolds numbers are in similar range.  In the $x_2$ direction, the cross-steam velocity itself is used for mean velocity excess since $\overline{U_{co,x_2}}=0$. The cross-stream mean velocity profile is also observed to show good agreement with the experimental results. 

%----------MEAN VELOCITY EXCESS---------------
\begin{figure}
\centering
\subfloat[]{%
\resizebox*{12.5cm}{!}{\includegraphics{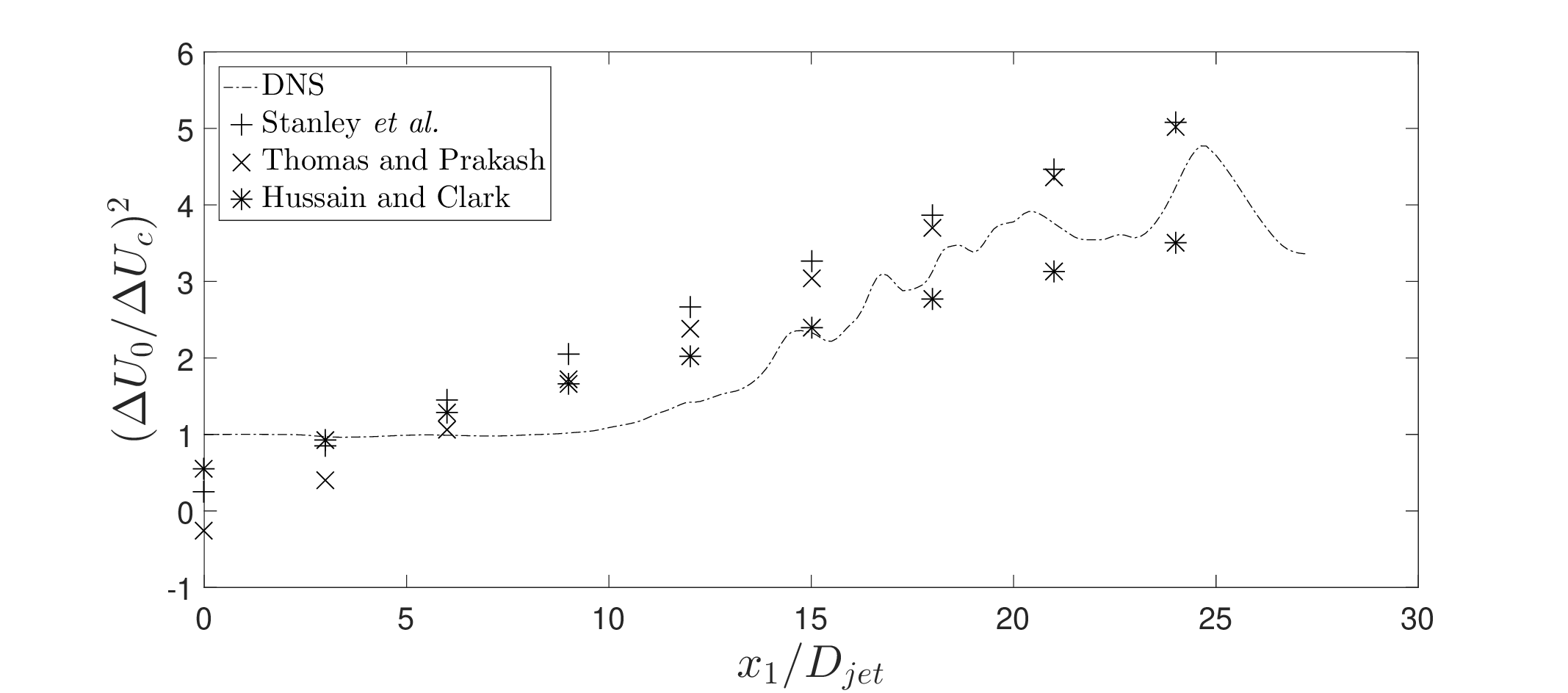}}}\hspace{5pt}
\subfloat[]{%
\resizebox*{12.5cm}{!}{\includegraphics{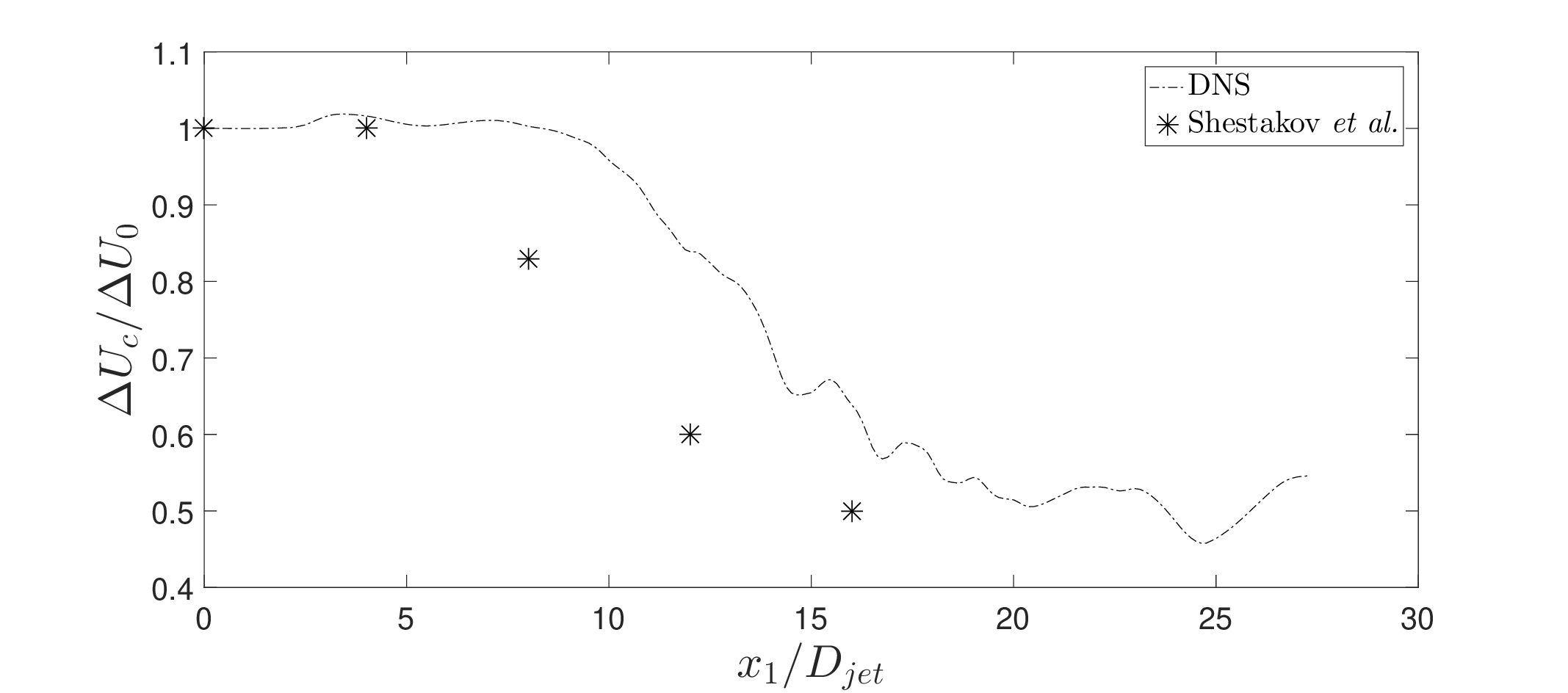}}}
\caption{Downstream growth of center-line mean velocity excess decay: a) Plane jet (experimental results vs current DNS), b) Slot jet (experimental results vs current DNS).}\label{cldecay}
\end{figure}

\begin{figure}
\centering
\subfloat[]{%
\resizebox*{12.5cm}{!}{\includegraphics{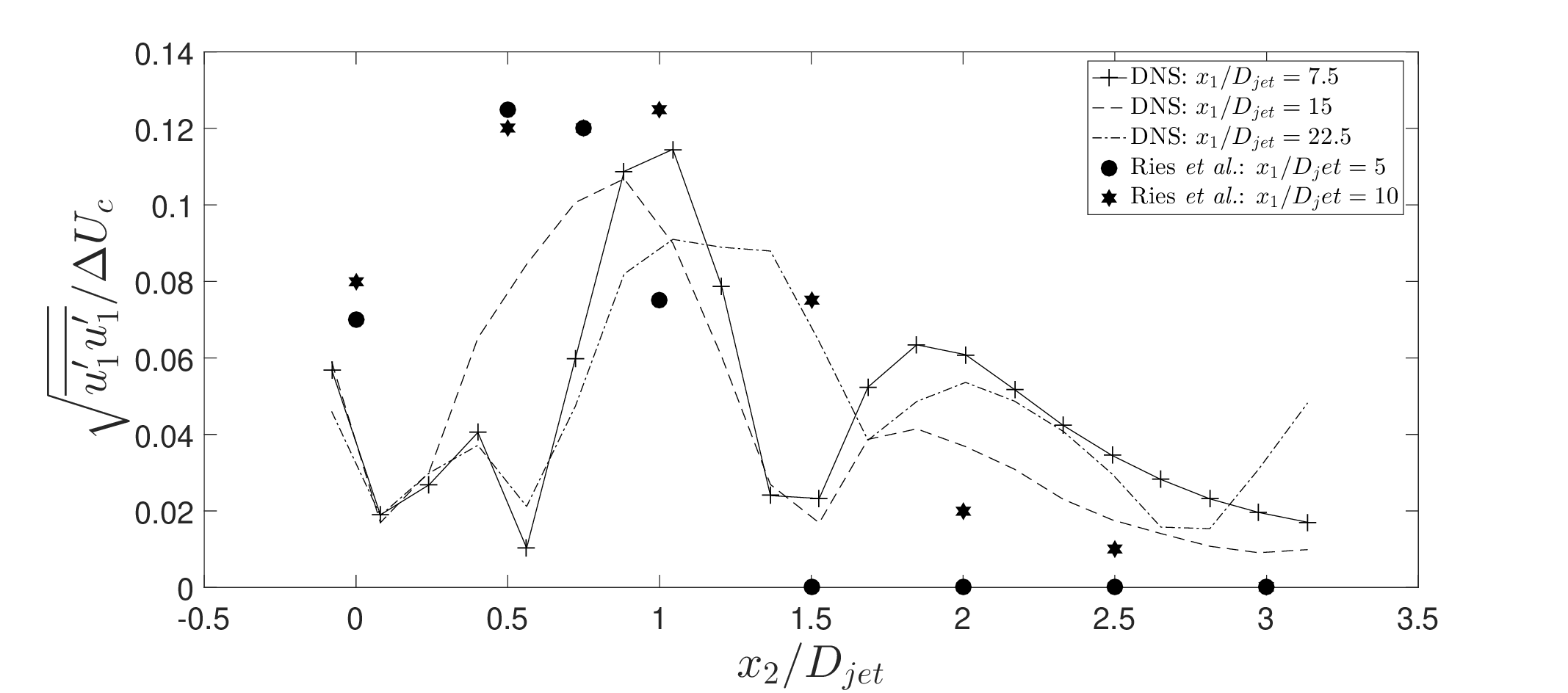}}}\hspace{5pt}
\subfloat[]{%
\resizebox*{12.5cm}{!}{\includegraphics{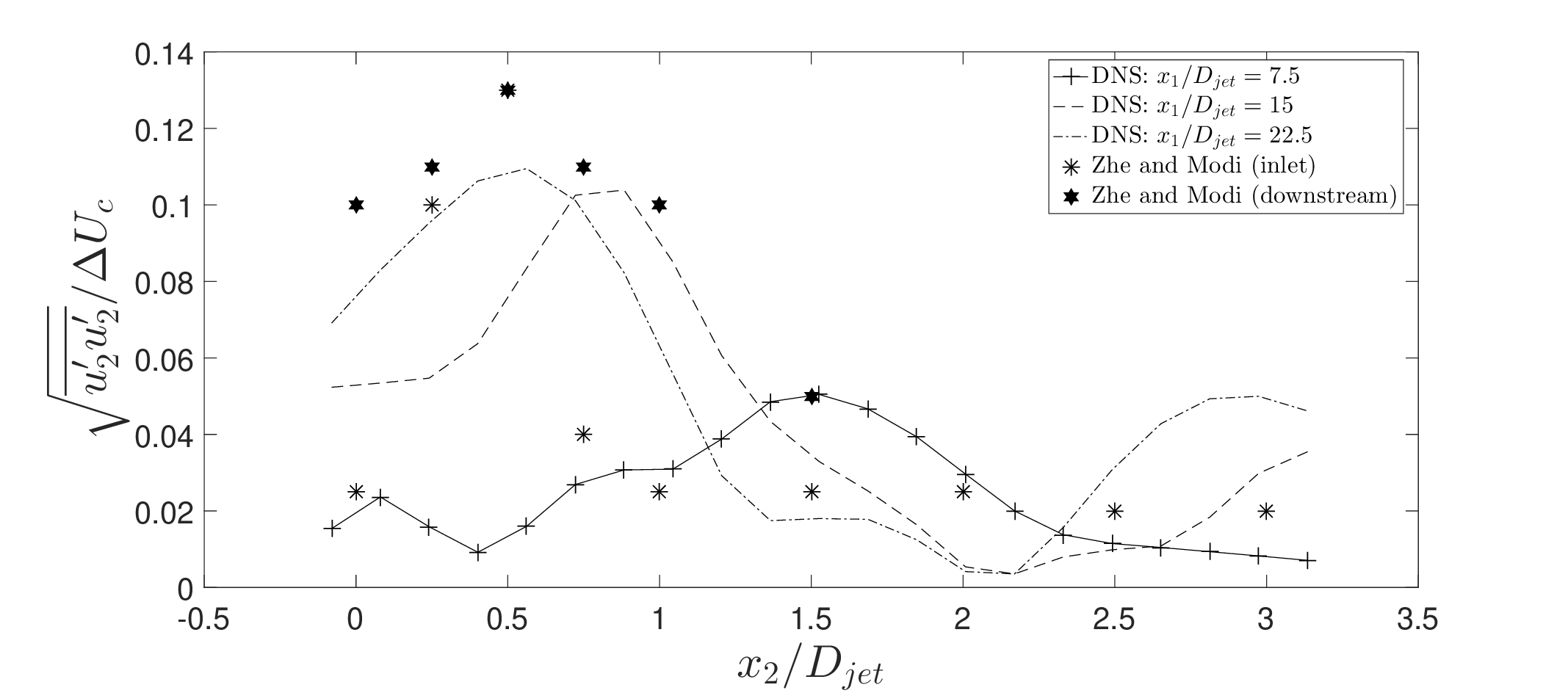}}}\hspace{5pt}
\subfloat[]{%
\resizebox*{12.5cm}{!}{\includegraphics{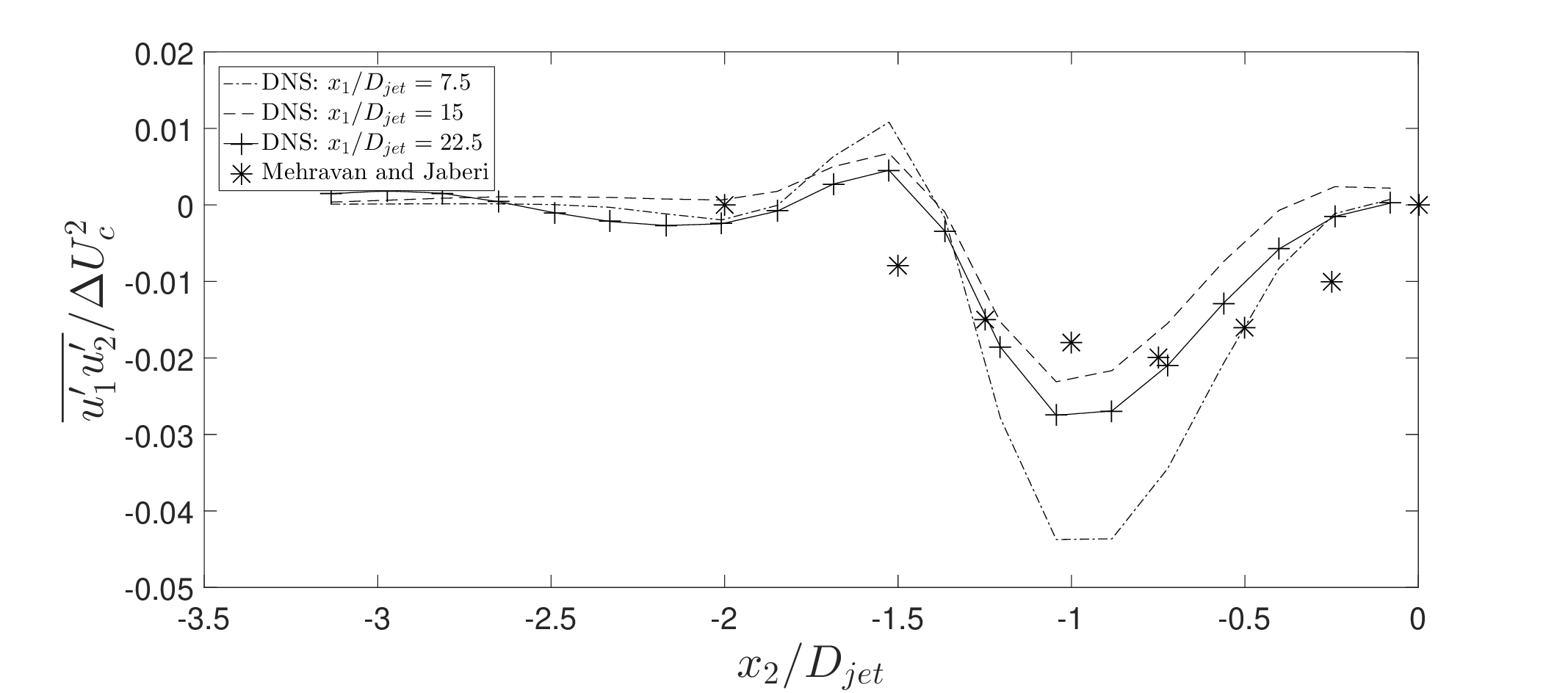}}}
\caption{Fluctuating r.m.s. velocity profile obtained from DNS compared with experimental data}
\label{rmsint}
\end{figure}

%---------------------------------------
The downstream growth of the decay of centerline velocity excess is presented in Fig. \ref{cldecay}, where $\Delta U_0$ denotes the difference between the mean centerline velocity and mean co-flow velocity at the inlet and $\Delta U_{c}$ denotes the difference between local mean centerline velocity and local mean co-flow velocity at downstream location. The centerline velocity decay is also observed to compare well with the experimental data of Shestakov \textit{et al.} \cite{slotjet}, Thomas and Prakash \cite{thomasprakash}, and Hussain and Clark \cite{hussainclark}. An inverse relationship between the downstream velocity decay and $x_1$ is obtained similar to the DNS results of Stanley \textit{et al.} \cite{stanleysarkar}. 
%------------- CL DECAY --------------------------------------
%---------------------------------------------
\begin{equation}
\Big[ \frac{\Delta U_0}{\Delta U_c} \Big]^2 = a_1 \Big[ \frac{x_1}{D_{jet}} + a_2 \Big],
\end{equation}
where $a_1$ and $a_2$ represent the slope and coefficient of curvefit, respectively. Table \ref{cldecayt} presents comparison of these values with results from the aforementioned experiments.
%%%%%%%%%%%%%%%%%%%%%% CL DECAY TABLE --------------------------------
\begin{table}
\tbl{Centerline mean velocity excess decay compared with several experiments and the physical parameters used in the experiments.}
{\begin{tabular}{lcccc} \toprule
Source                   & $a_1$    & $a_2$    & $Re_0$ & $h/\theta$ \\ \midrule
DNS                                              & 0.17  & 4.58  & 850    & 20         \\
Stanley \textit{et al.} & 0.201 & 1.23  & 3000   & 20         \\
Thomas and Prakash        & 0.220 & -1.20 & 8000   & 67         \\
Gutmark and Wygnanski           & 0.189 & -4.72 & 30000  & -          \\
Hussain and Clark          & 0.123 & 4.47  & 32552  & 182        \\
\bottomrule
\end{tabular}}
\label{cldecayt}
\end{table}
%-------------------------------------------------------------
The root mean square (r.m.s.) of the velocity fluctuations measured from the DNS are compared against DNS of turbulent round jets by Ries \textit{et al.} \cite{riesetal}, experiments on turbulent slot jets by Zhe and Modi \cite{zhemodi}, experiments on turbulent plane jets by Khayrullina \textit{et al.} \cite{khayru} and presented in Fig. \ref{rmsint}. The r.m.s. of velocity fluctuations show good agreement to the experimental data.
%-----------RMS--------------
%---------------------------------------------
%%%%%%%%%%%%%%%%%%%%%%
\subsection{FMDF}
The FMDF of mixture fraction for varying pressures, Reynolds numbers and diffusion models obtained from the DNS near the jet centerline. The FMDF is measured after the flow has reached a statistically stationary state. The shapes of the FMDF describe the distribution of the instantaneous SGS fluctuations. When the mixture fraction is highly segregated, the shape of the FMDF is bimodal, else it is unimodal indicating high concentrations of either fuel or oxidizer \cite{fdfexp,fdfexp2}. In highly dispersed flows, multimodal shapes may also be observed. In non-premixed flows, the regions close to the nozzle along the jet centerline and freestream typically have high concentrations of fuel and oxidizer, respectively.

\begin{figure}
\centering
\includegraphics[width=1\linewidth]{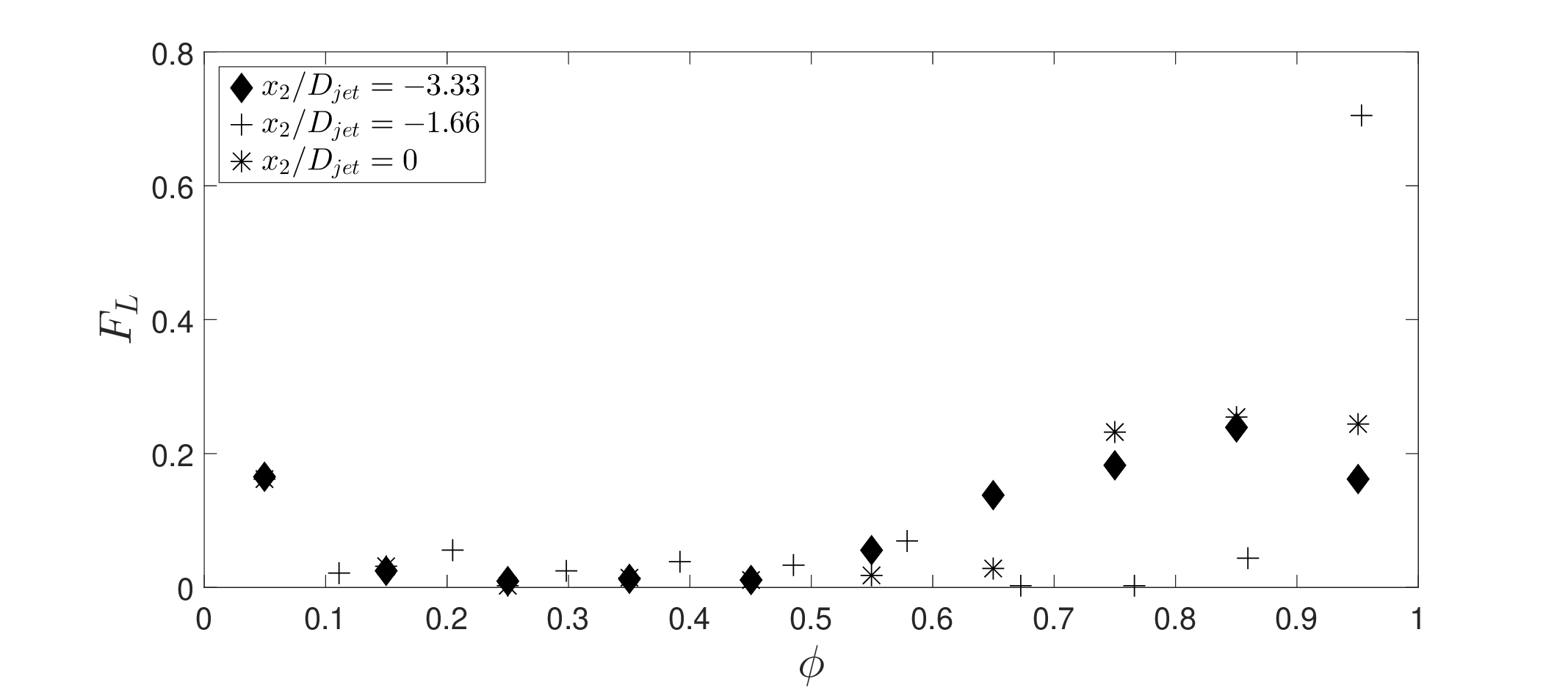}
\caption{Filtered mass density function of mixture fraction measured from the DNS at $P_0=1 atm$, $Re_0=850$, $Le \neq 1$ with the generalized diffusion model. Data extracted at $x_1=24 D_{jet}$ with a filter size of $\Delta_f=4 \Delta x_1$}
\label{fmdfmixf3}
\end{figure}

\begin{figure}
\centering
\includegraphics[width=1\linewidth]{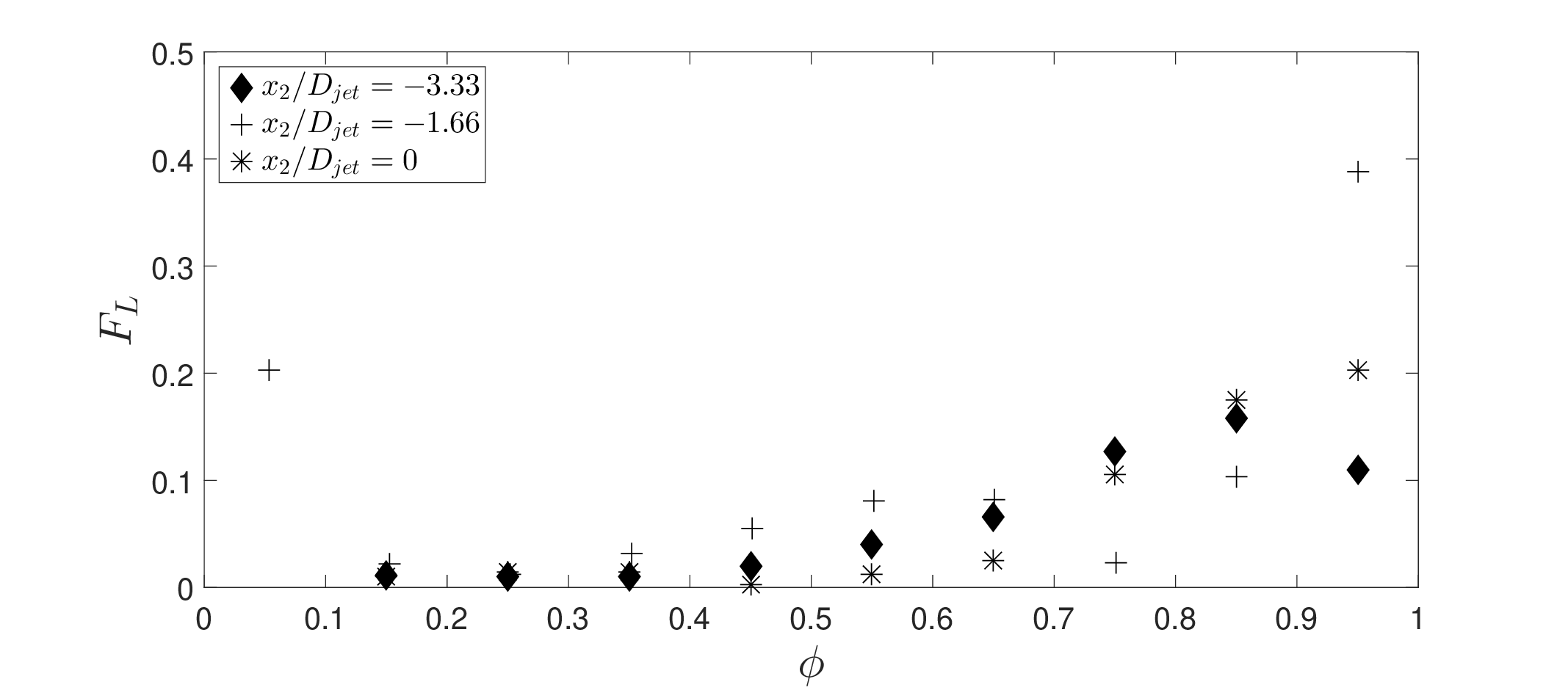}
\caption{Filtered mass density function of mixture fraction measured from the DNS at $P_0=1 atm$, $Re_0=1300$, $Le \neq 1$ with the generalized diffusion model. Data extracted at $x_1=24 D_{jet}$ with a filter size of $\Delta_f=4 \Delta x_1$}
\label{fmdfmixf5}
\end{figure}

In simulations at atmospheric pressures with the standard Fickian diffusion model, the FMDF is observed to be predominantly unimodal at the jet centerline. The FMDF at $P_0=100 atm$ with the Fickian diffusion model is similar to that at $P_0=1 atm$ indicating no significant change with pressure. However, at atmospheric pressures with the generalized diffusion model, the FMDF is bimodal as seen in Fig. \ref{fmdfmixf3}. The bimodal shape of the FMDF at $P_0=1 atm$ with the generalized diffusion model is seen to become more segregated with increase in Reynolds number (Fig. \ref{fmdfmixf5}). With increase in pressure, the FMDF becomes multimodal. This indicates intensification of mixing and increase in rate of diffusion with increase in pressure. With the generalized diffusion model [Eq. \ref{mass flux}], a significant change in the FMDF is observed; especially at $P_0>1 atm$. The generalized diffusion model takes into account the diffusion due to concentration and pressure gradients which become very significant at large pressures.
%--------------- FMDF ---------------------------
\begin{figure}
\centering
\includegraphics[width=1\linewidth]{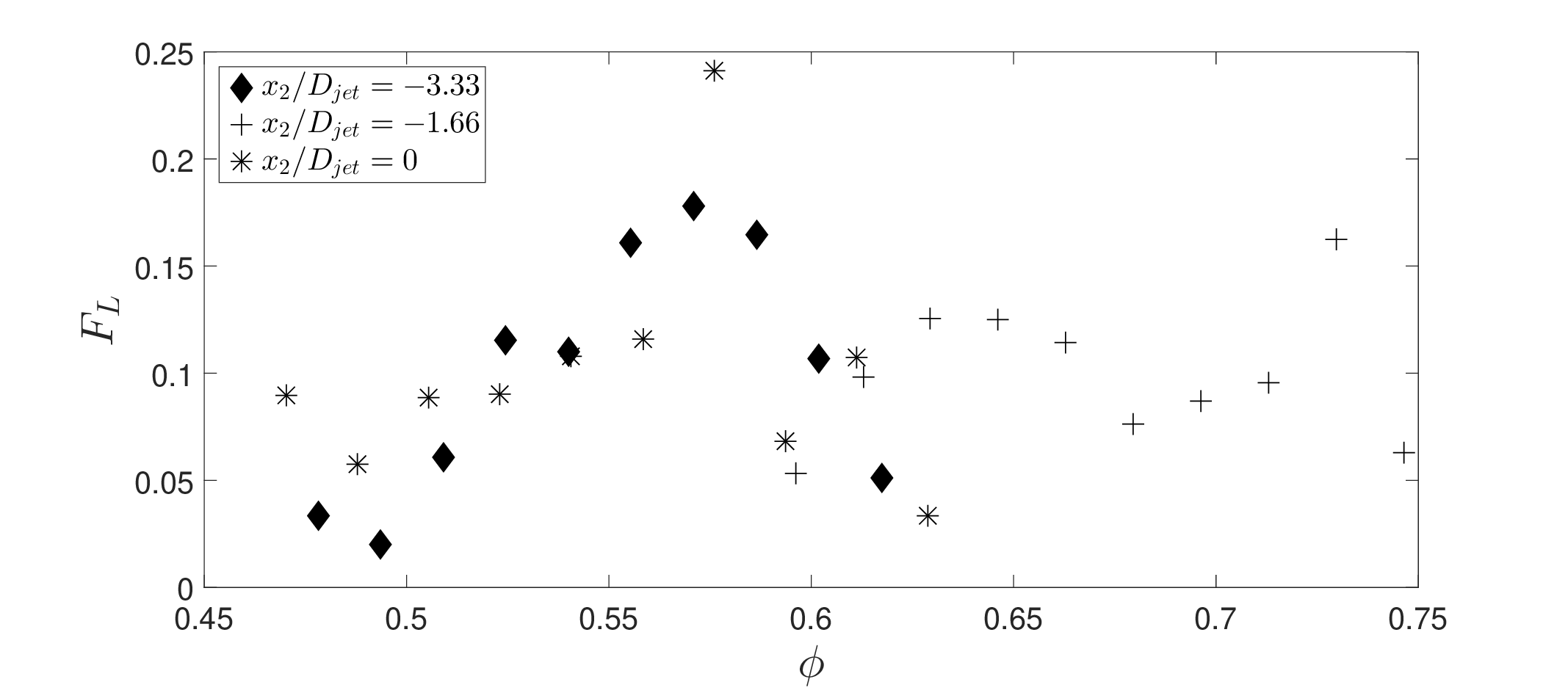}
\caption{Filtered mass density function of mixture fraction measured from the DNS at $P_0=1 atm$, $Re_0=850$, $Le=1$ with the standard Fickian diffusion model. Data extracted at $x_1=24 D_{jet}$ with a filter size of $\Delta_f=4 \Delta x_1$}
\label{fmdfmixf1}
\end{figure}
%-------------------------------------
\begin{figure}
\centering
\includegraphics[width=1\linewidth]{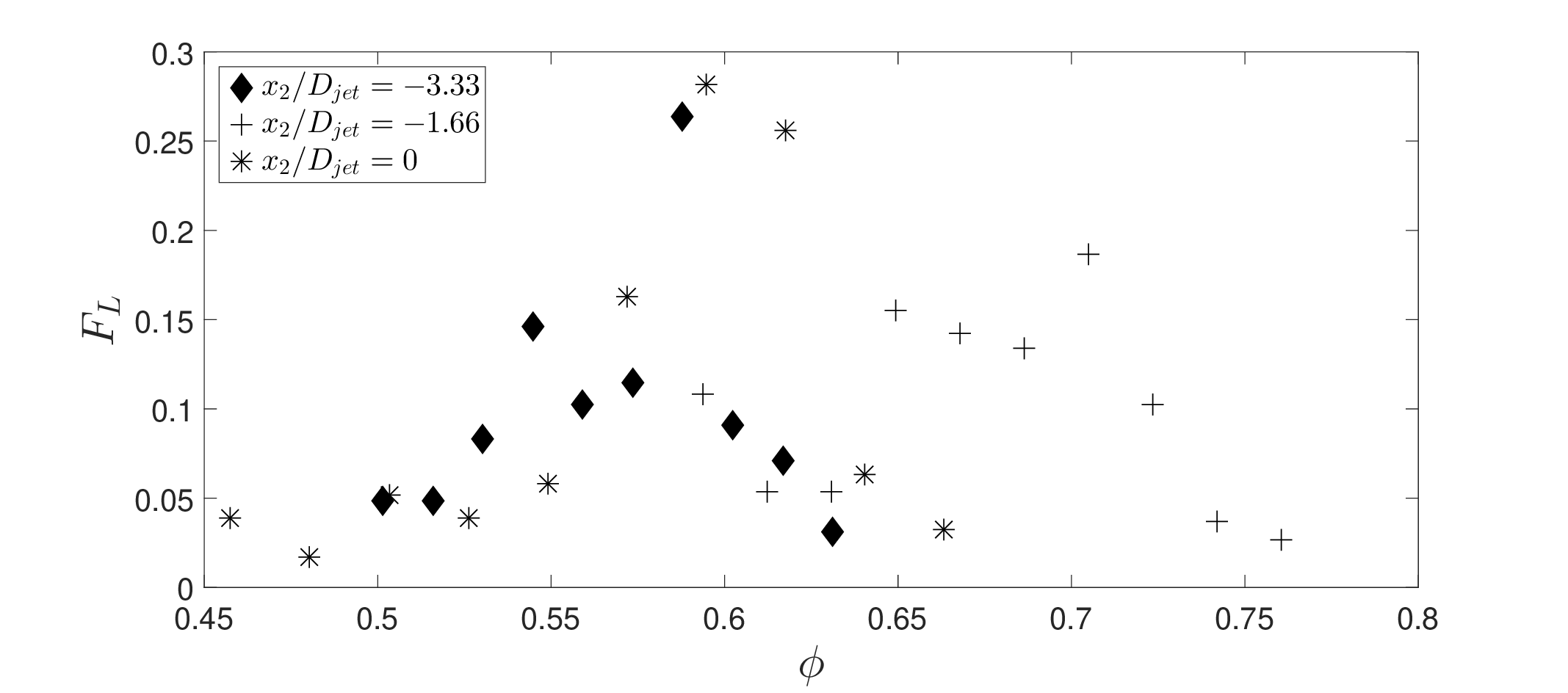}
\caption{Filtered mass density function of mixture fraction measured from the DNS at $P_0=100 atm$, $Re_0=850$, $Le=1$ with the standard Fickian diffusion model. Data extracted at $x_1=24 D_{jet}$ with a filter size of $\Delta_f=4 \Delta x_1$}
\label{fmdfmixf2}
\end{figure}
%-------------------------------------------------------------
%-------------------------------------------------------------
%-------------------------------------------------------------
\subsection{CSD}
\begin{figure}
\centering
{\resizebox*{12.5cm}{!}{\includegraphics{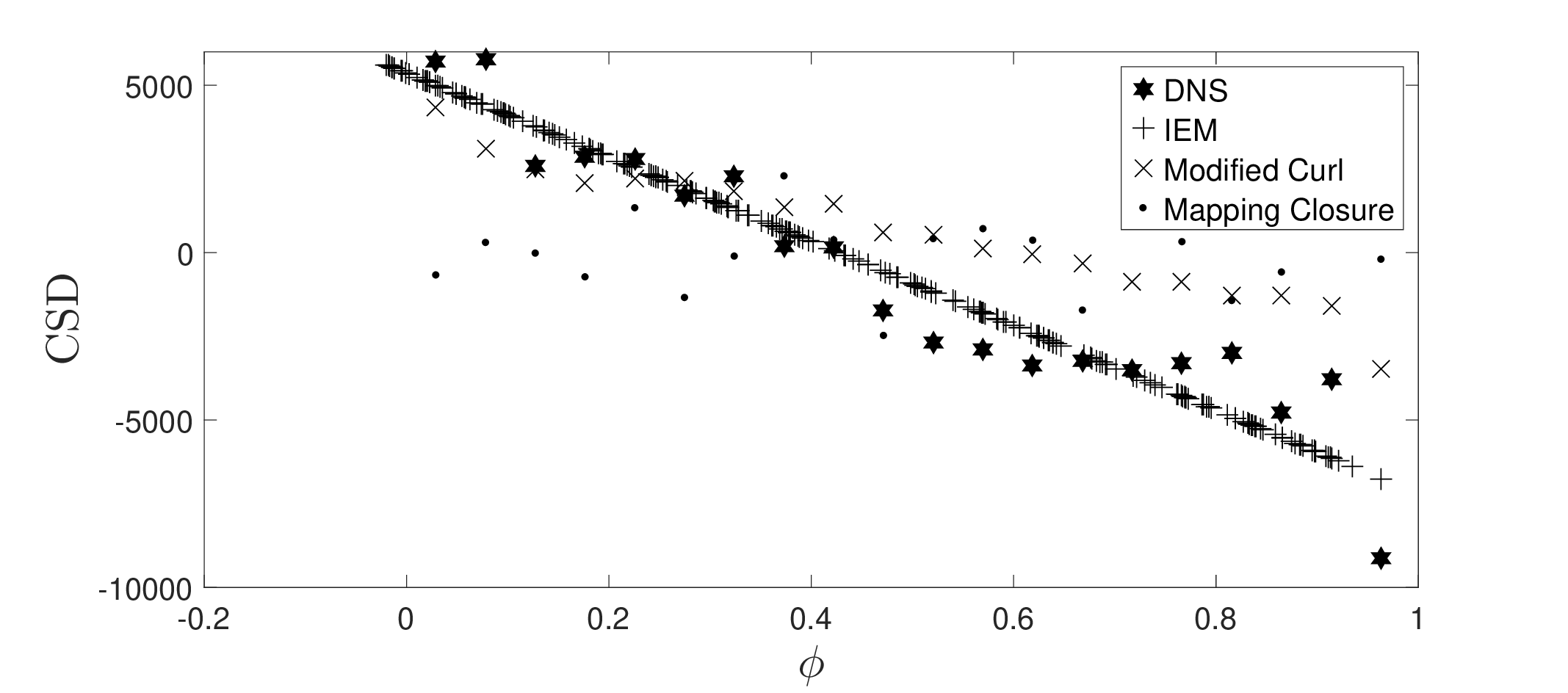}}}
\caption{CSD $\big(\big\langle \frac{1}{\rho} \frac{\partial J_{\phi,j}}{\partial x_j} \big| \phi \big\rangle \big)$ modeled by the IEM, MC and MAPPING models. CSD obtained from the DNS of heptane-air mixture at $P_0=1 atm$, $Re_0=850$, $Le=1$ with ideal gas equation of state and the Fickian diffusion model. $\Omega_{m,calculated}=1.26 \times 10^4$.}
\label{csdm1FNI850}
\end{figure}

\begin{figure}
\centering
\subfloat[]{%
\resizebox*{12.5cm}{!}{\includegraphics{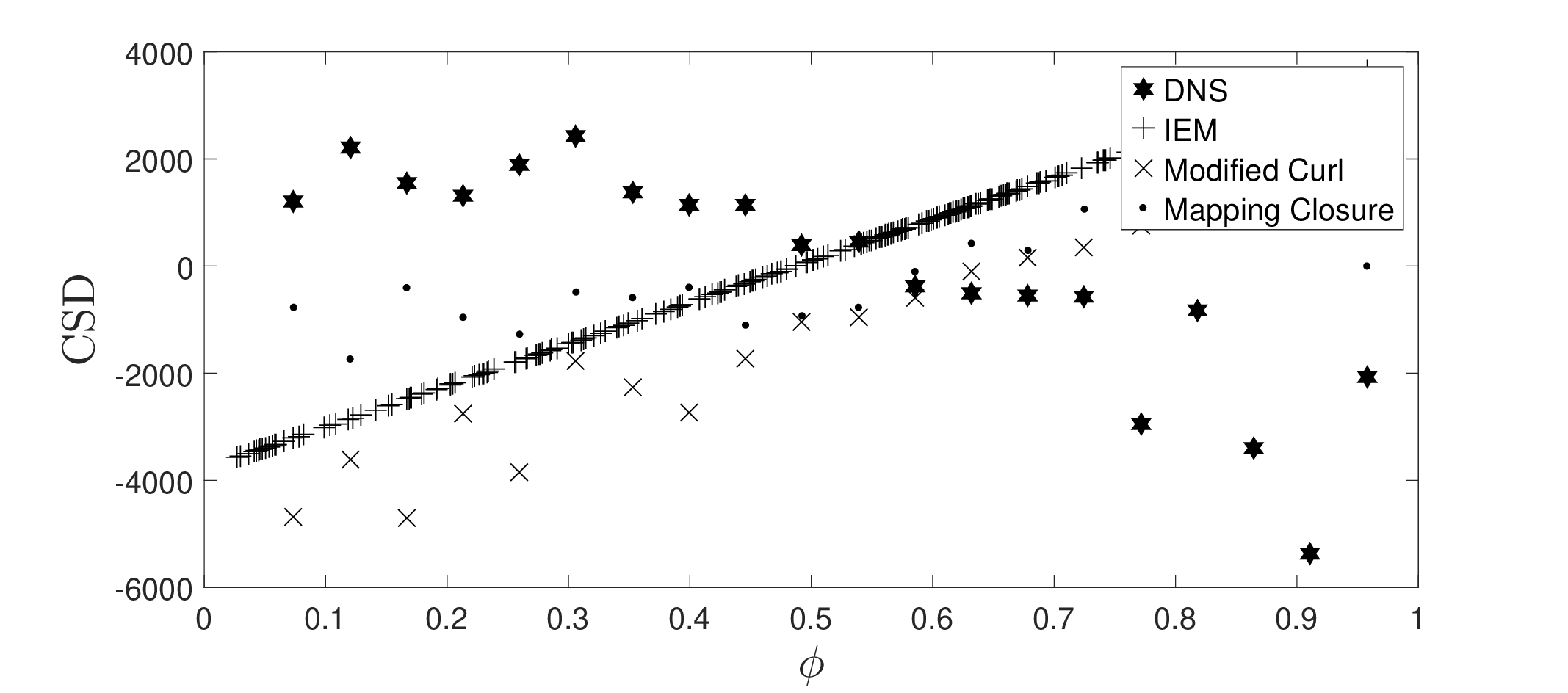}}}\hspace{5pt}
\subfloat[]{%
\resizebox*{12.5cm}{!}{\includegraphics{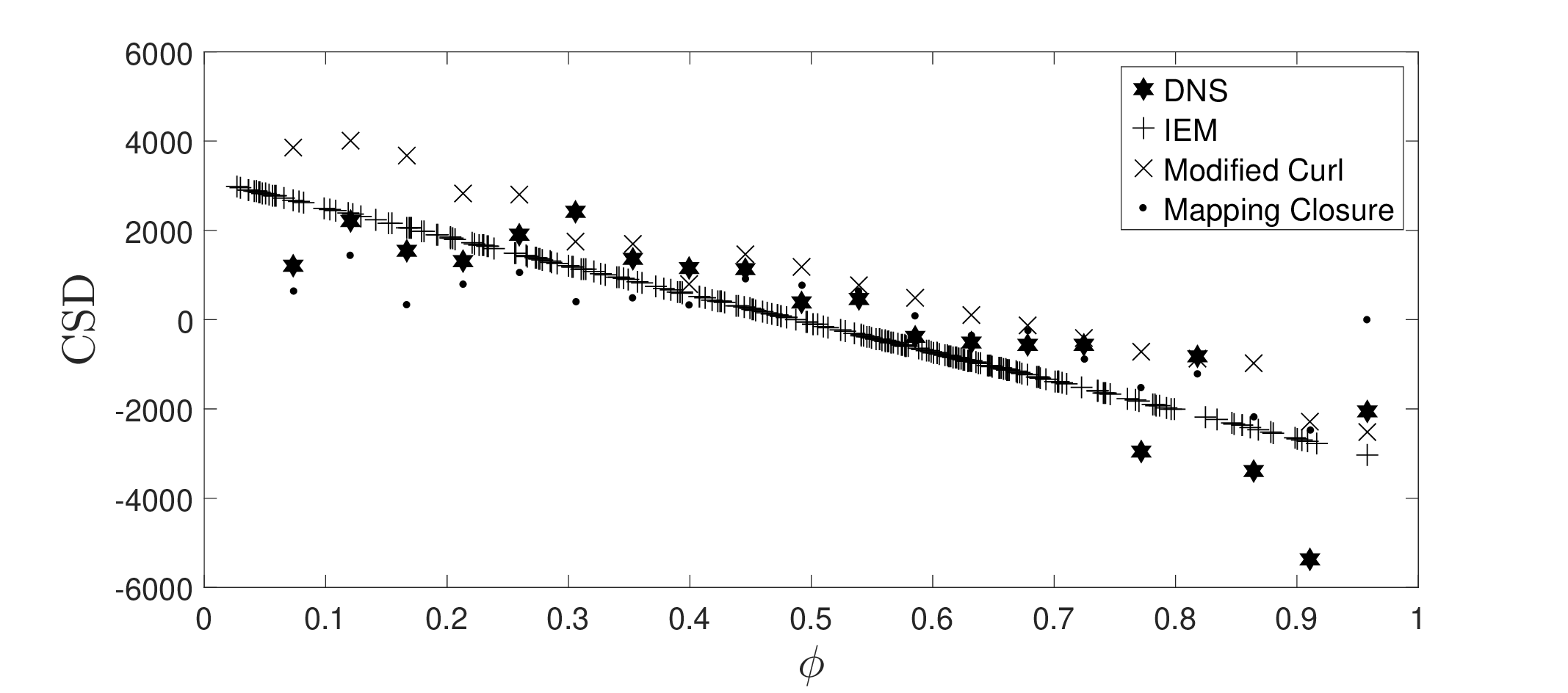}}}
\caption{CSD $\big(\big\langle \frac{1}{\rho} \frac{\partial J_{\phi,j}}{\partial x_j} \big| \phi \big\rangle \big)$ modeled by the IEM, MC and MAPPING models. CSD obtained from the DNS of heptane-air mixture at $P_0=1 atm$, $Re_0=850$, $Le \neq 1$ with real gas equation of state and the generalized diffusion model: a) $\Omega_{m,calculated}=-7.76 \times 10^3$, b) $\Omega_{m,corrected}=6.45 \times 10^3$.}
\label{csdm1GNR850}
\end{figure}

\begin{figure}
\centering
\subfloat[]{%
\resizebox*{12.5cm}{!}{\includegraphics{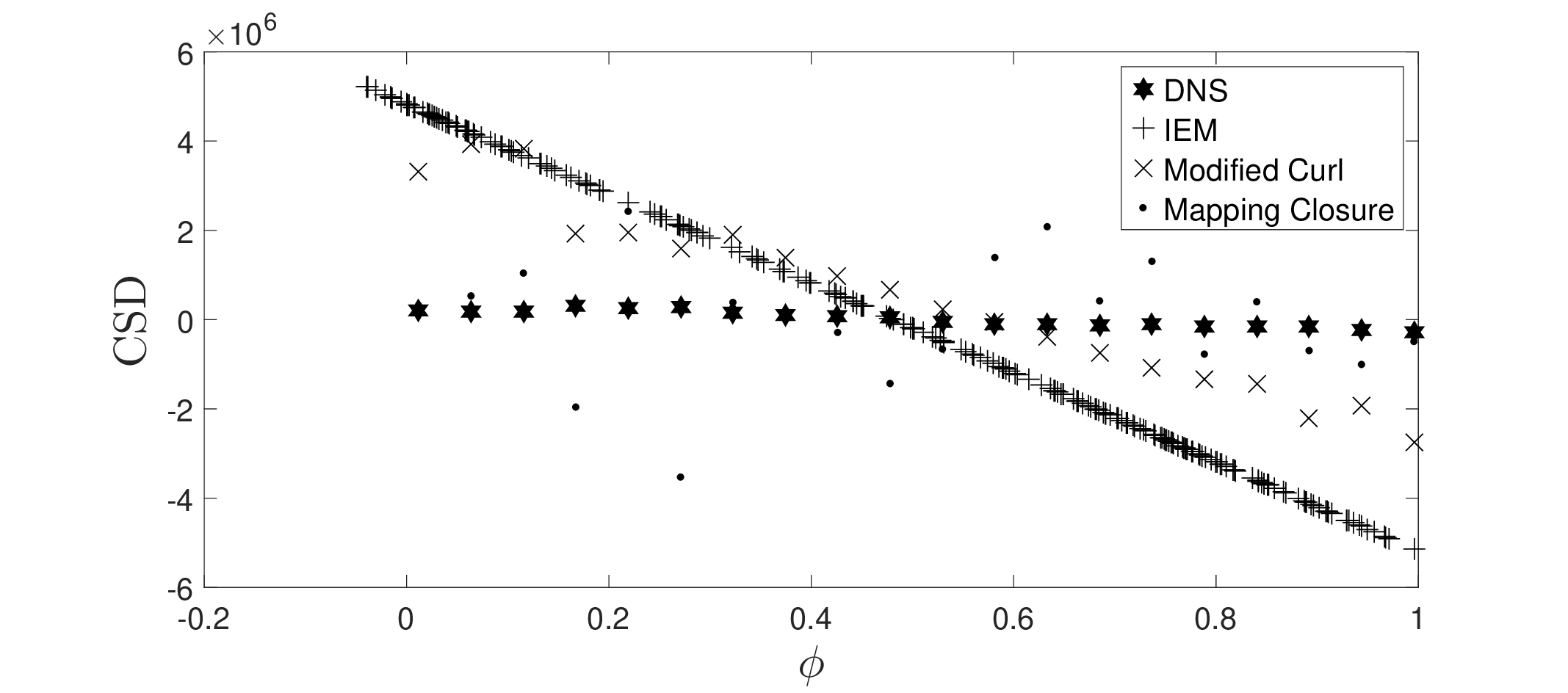}}}\hspace{5pt}
\subfloat[]{%
\resizebox*{12.5cm}{!}{\includegraphics{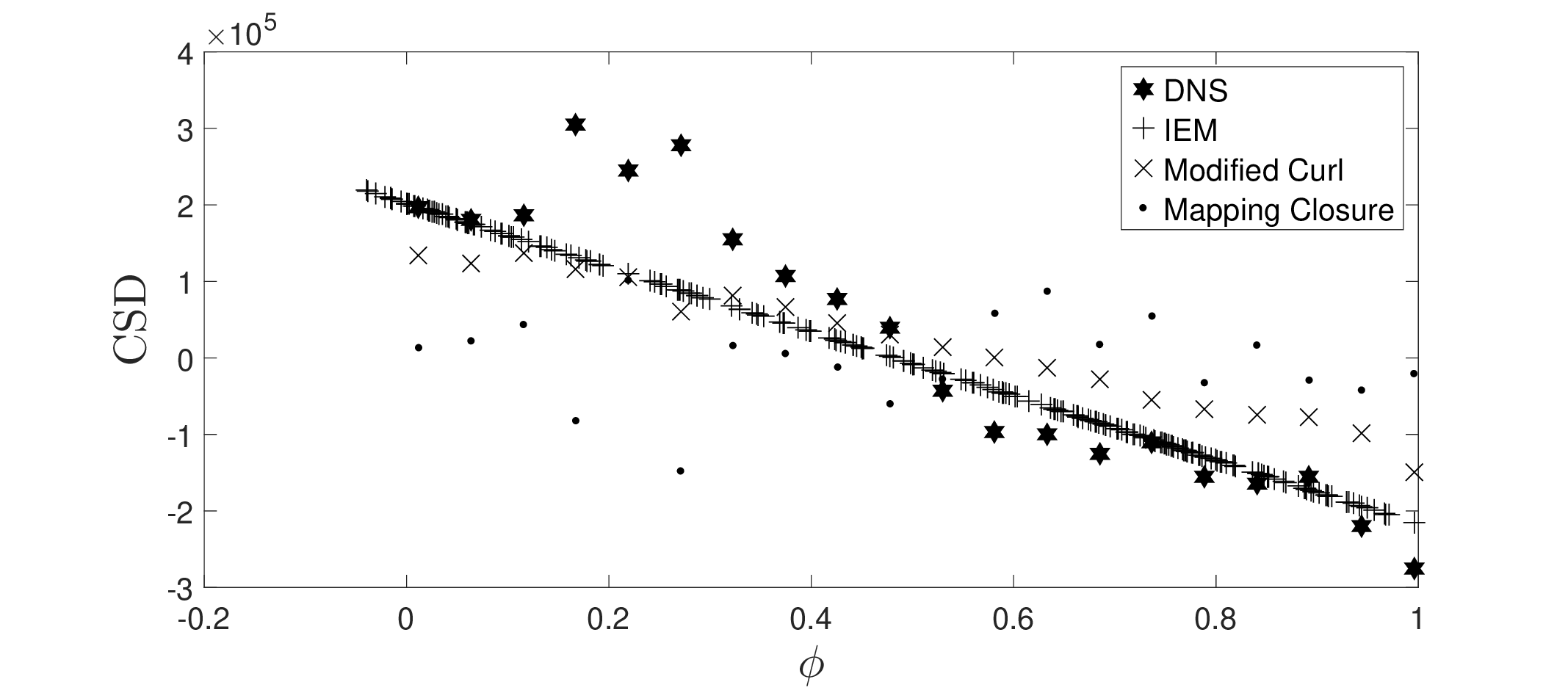}}}
\caption{CSD $\big(\big\langle \frac{1}{\rho} \frac{\partial J_{\phi,j}}{\partial x_j} \big| \phi \big\rangle \big)$ modeled by the IEM, MC and MAPPING models. CSD obtained from the DNS of heptane-air mixture at $P_0=35 atm$, $Re_0=850$, $Le \neq 1$ with real gas equation of state and the generalized diffusion model: a) $\Omega_{m,calculated}=1.34 \times 10^7$, b) $\Omega_{m,corrected}=5.59 \times 10^5$.}
\label{csdm35GNR850}
\end{figure}

\begin{figure}
\centering
\subfloat[]{%
\resizebox*{12.5cm}{!}{\includegraphics{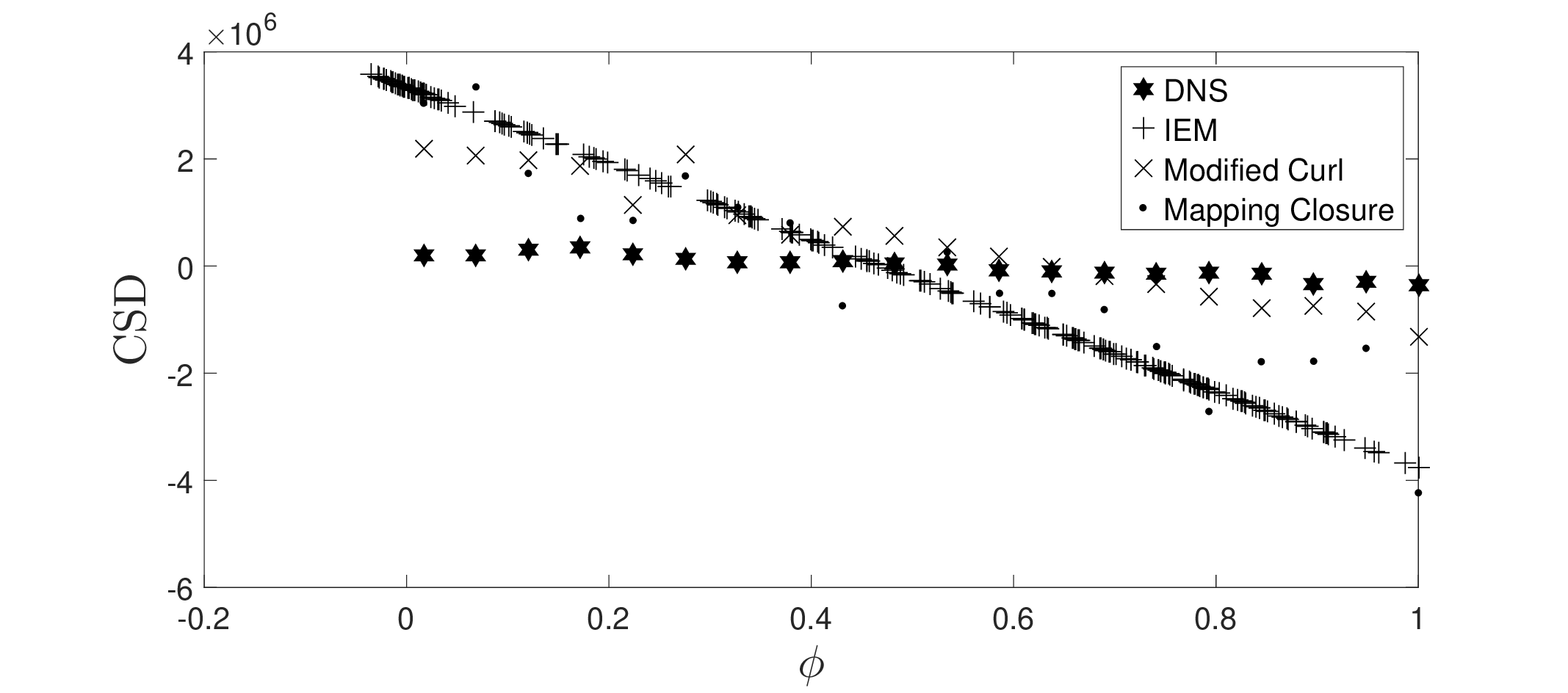}}}\hspace{5pt}
\subfloat[]{%
\resizebox*{12.5cm}{!}{\includegraphics{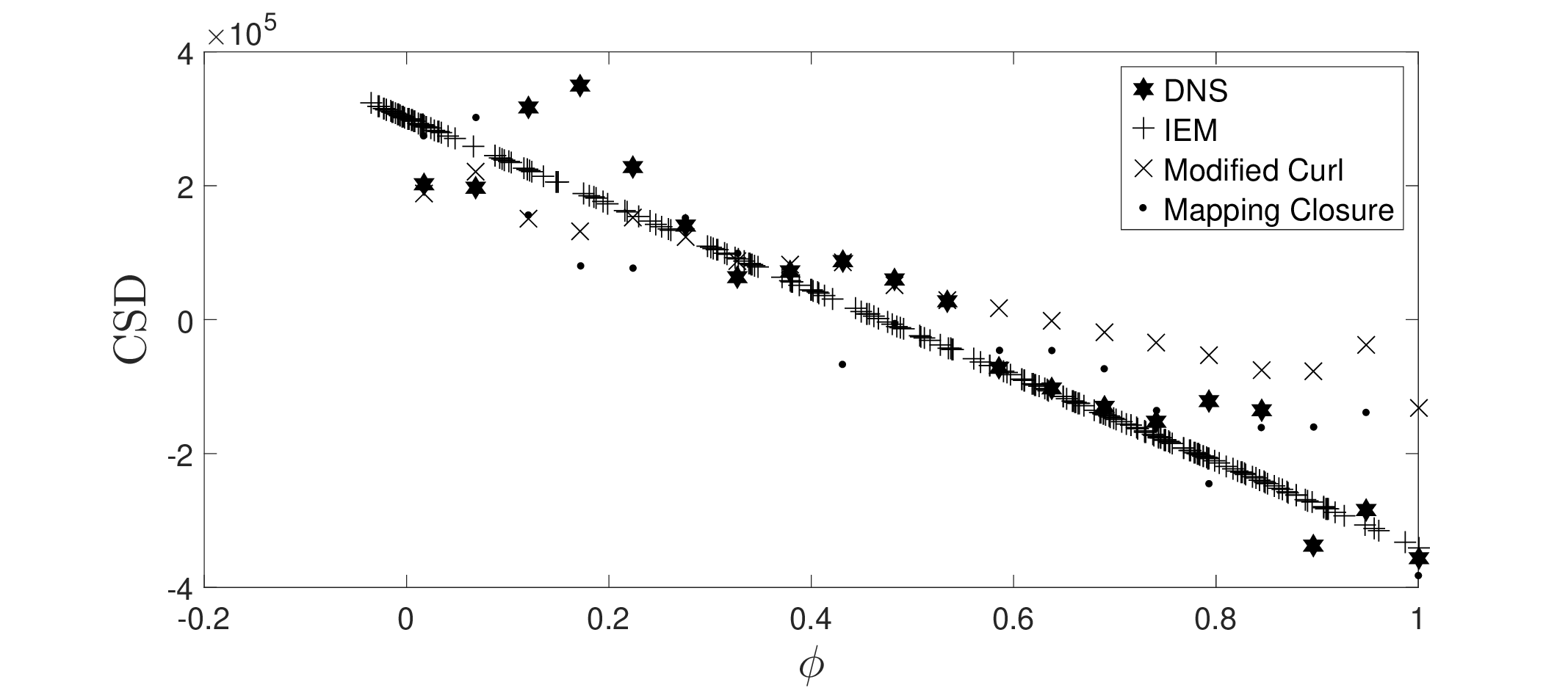}}}
\caption{CSD $\big(\big\langle \frac{1}{\rho} \frac{\partial J_{\phi,j}}{\partial x_j} \big| \phi \big\rangle \big)$ modeled by the IEM, MC and MAPPING models. CSD obtained from the DNS of heptane-air mixture at $P_0=100 atm$, $Re_0=850$, $Le \neq 1$ with real gas equation of state and the generalized diffusion model: a) $\Omega_{m,calculated}=7.10 \times 10^6$, b) $\Omega_{m,corrected}=6.41 \times 10^5$.}
\label{csdm100GNR850}
\end{figure}

\begin{figure}
\centering
\subfloat[]{%
\resizebox*{12.5cm}{!}{\includegraphics{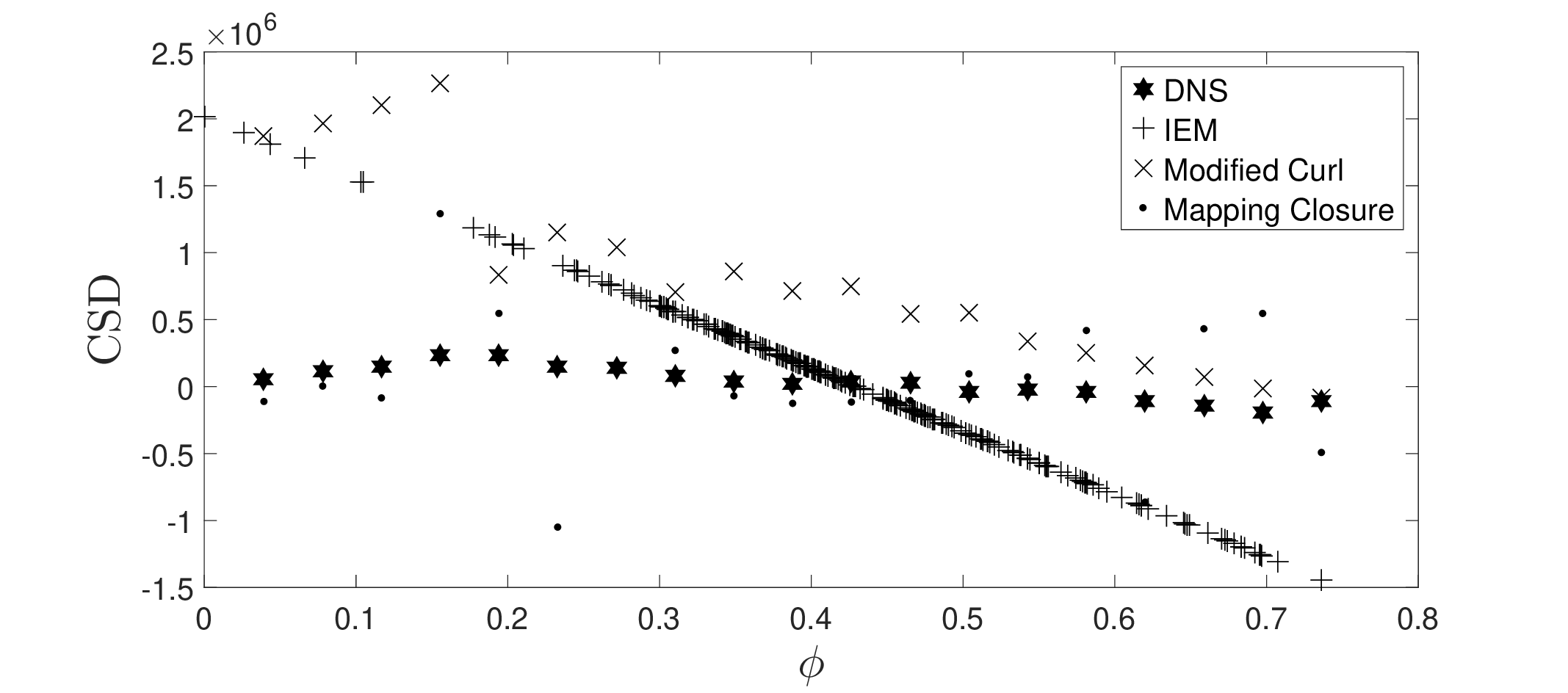}}}\hspace{5pt}
\subfloat[]{%
\resizebox*{12.5cm}{!}{\includegraphics{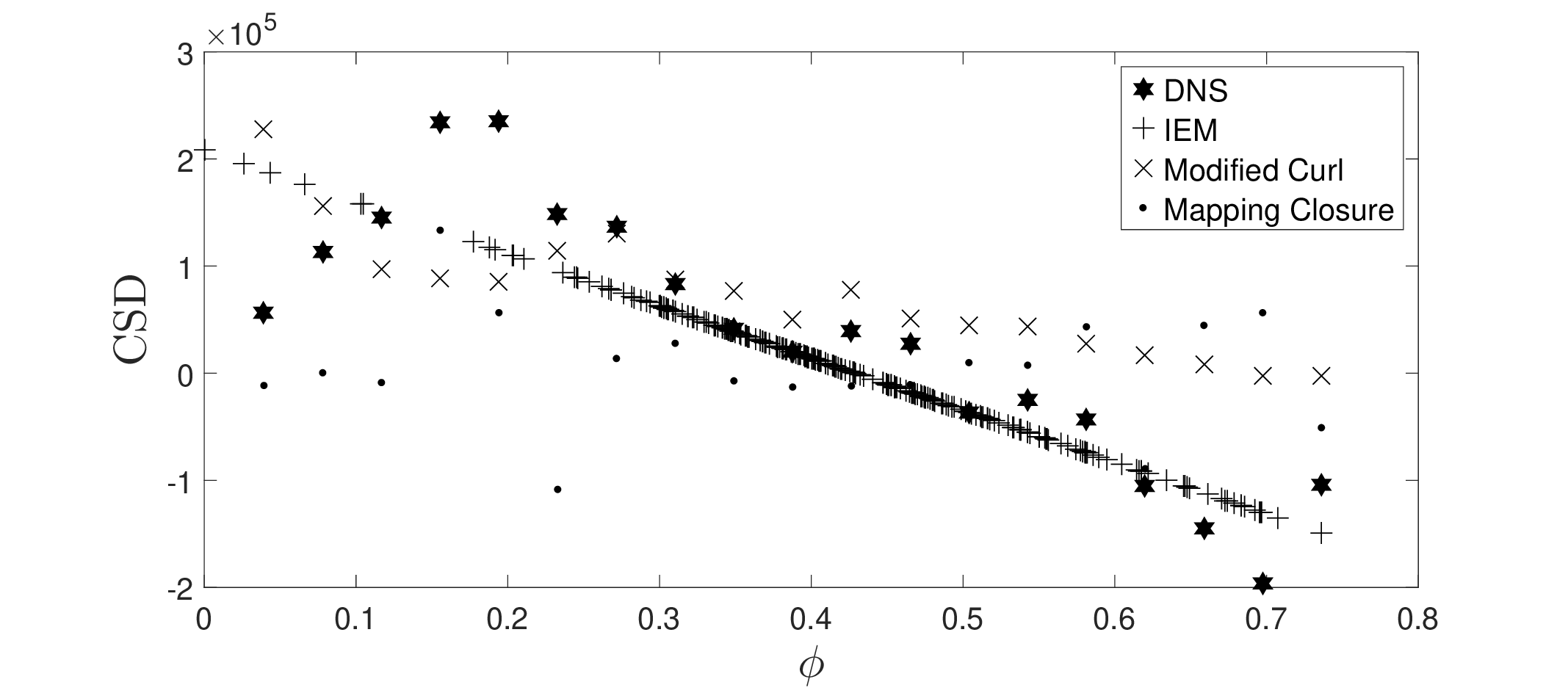}}}
\caption{CSD $\big(\big\langle \frac{1}{\rho} \frac{\partial J_{\phi,j}}{\partial x_j} \big| \phi \big\rangle \big)$ modeled by the IEM, MC and MAPPING models. CSD obtained from the DNS of heptane-air mixture at $P_0=100 atm$, $Re_0=1300$, $Le \neq 1$ with real gas equation of state and the generalized diffusion model: a) $\Omega_{m,calculated}=4.7 \times 10^6$, b) $\Omega_{m,corrected}=4.8 \times 10^5$.}
\label{csdm100GNR1300}
\end{figure}

\begin{figure}
\centering
\subfloat[]{%
\resizebox*{12.5cm}{!}{\includegraphics{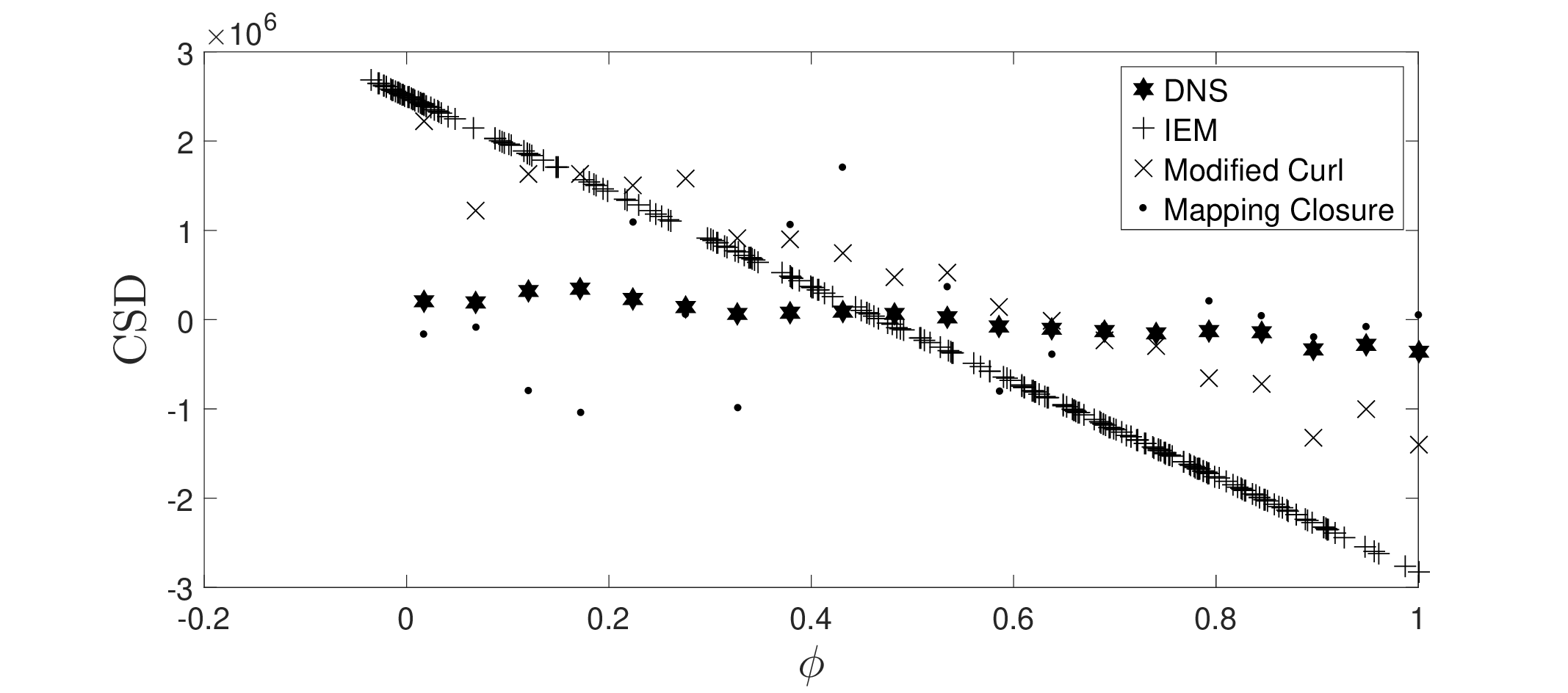}}}\hspace{5pt}
\subfloat[]{%
\resizebox*{12.5cm}{!}{\includegraphics{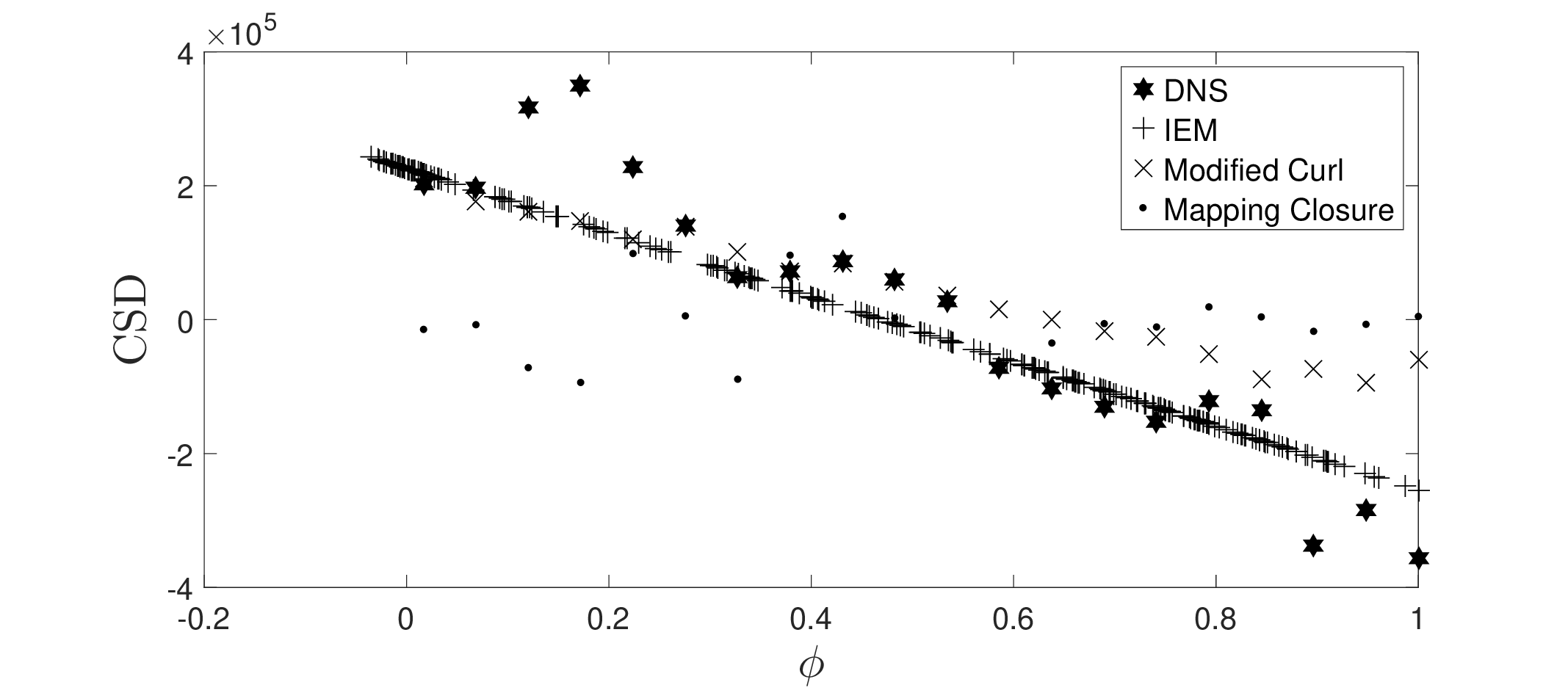}}}
\caption{CSD $\big(\big\langle \frac{1}{\rho} \frac{\partial J_{\phi,j}}{\partial x_j} \big| \phi \big\rangle \big)$ modeled by the IEM, MC and MAPPING models. CSD obtained from the DNS of heptane-oxygen mixture at $P_0=100 atm$, $Re_0=850$, $Le \neq 1$ with real gas equation of state and the generalized diffusion model: a) $\Omega_{m,calculated}=7.10 \times 10^6$, b) $\Omega_{m,corrected}=6.41 \times 10^5$.}
\label{csdm100GNR850_hepo}
\end{figure}

The CSD measured from the DNS at pressures of $1 atm$, $35 atm$ and $100 atm$ with the generalized and the simplified Fickian diffusion model [Eq. \ref{standard fickian}] are compared against the mixing models and presented in Figs. \ref{csdm1FNI850}-\ref{csdm100GNR850_hepo}. The CSD is stochastic in nature but the variation in shape and magnitude of the term with respect to time is observed to be relatively small. The absolute magnitude of the CSD is observed to increase non-linearly with increase in ambient pressure. A discrete set of scalar diffusion values values within a filter volume are conditioned with the mixture fraction. Hence, the shape of the CSD is discontinuous.
% CSD MODELS
%------------1--------------
%-------------------------------
%-----------2---------------
%-----------3----------------
%-------------------------------
%-----------4----------------
%-------------------------------
%-----------5----------------
%-------------------------------
%-----------6----------------
%-------------------------------
\subsection{Comparison of CSD and Mixing models}
\subsubsection{Interaction by Exchange of Mean}
The parameters for mixing frequency ($\Omega_m$) of the IEM model described in Section \ref{iem_intro} are chosen as $C_{\Omega} = 4$, $C_R = 0.013$, $\Delta_G=12$ and $Sc_t = 0.7$. The parameters are obtained from the works of Jaberi \textit{et al.} \cite{fmdfintro}. The IEM model fits the CSD linearly. The trends predicted by the model are observed to be exact for flows at atmospheric pressures with both ideal and real gas equations of state, and with the standard Fickian and generalized diffusion models. However, at $P_0=100 atm$ with the Fickian diffusion model, the IEM model over-predicts the CSD by a factor of 10. Similar trends are observed in flows at $P_0 > 1 atm$ with the generalized diffusion model. The reason for the difference between the DNS data and model is determined to be the mixing frequency ($\Omega_m$). The mixing frequency is observed to be over-predicted by a factor of 10 or larger in flows at high pressures with the generalized diffusion model and under-predicted by a factor of 10 in flows at high pressure with the Fickian diffusion model. The least-squares curve fit method is used to calibrate the model parameters to fit to the DNS data and to determine the corrected mixing frequencies. The calculated and corrected mixing frequencies for various test cases are presented in Table \ref{mixfreqcomp}.

\begin{figure}
\centering
{%
\resizebox*{12.5cm}{!}{\includegraphics{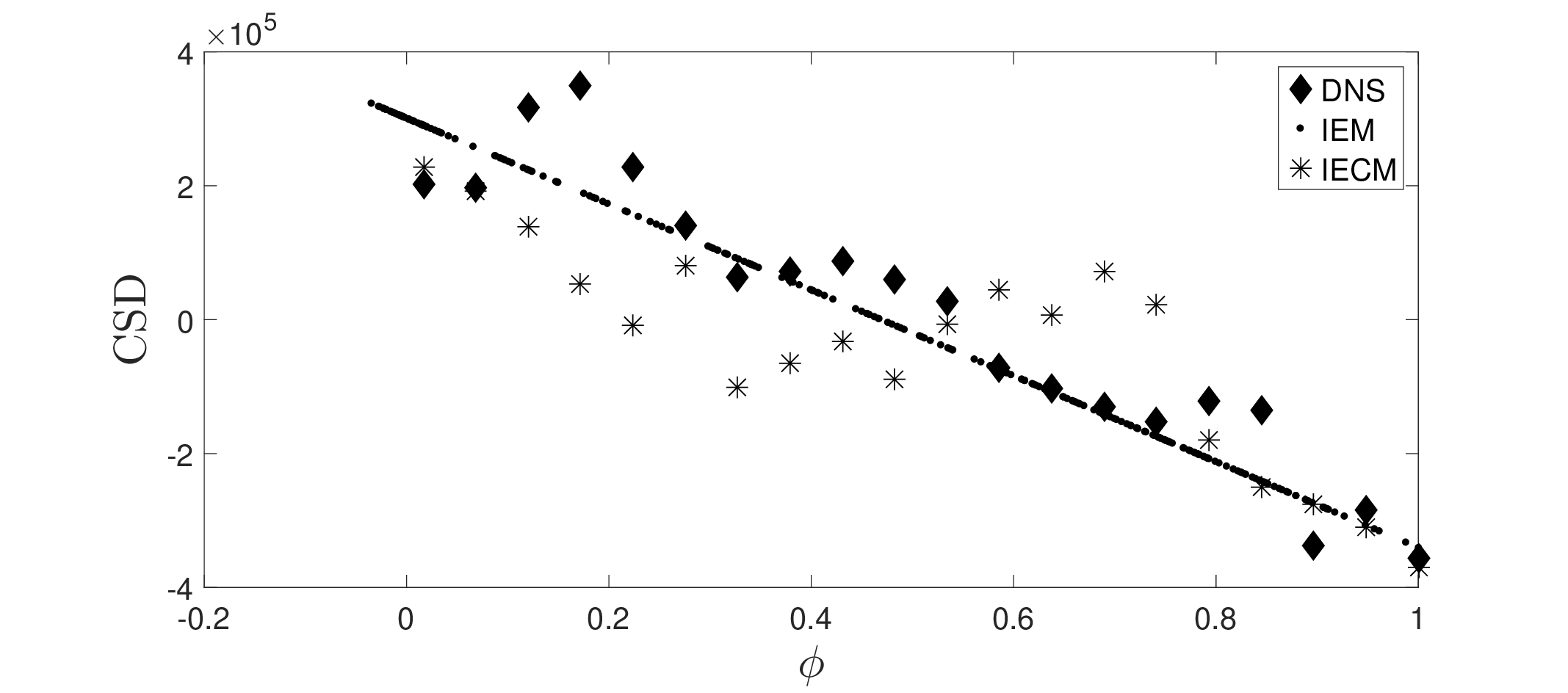}}}\hspace{5pt}
\caption{CSD $\big(\big\langle \frac{1}{\rho} \frac{\partial J_{\phi,j}}{\partial x_j} \big| \phi \big\rangle \big)$ modeled by the IECM model. CSD obtained from the DNS of heptane-air mixture at $P_0=100 atm$, $Re_0=850$, $Le \neq 1$ with real gas equation of state and the generalized diffusion model. The mixing frequency for this model is same as that of the IEM model.}
\label{csdm100GNR850_iecm}
\end{figure}

\begin{figure}
\centering
{%
\resizebox*{12.5cm}{!}{\includegraphics{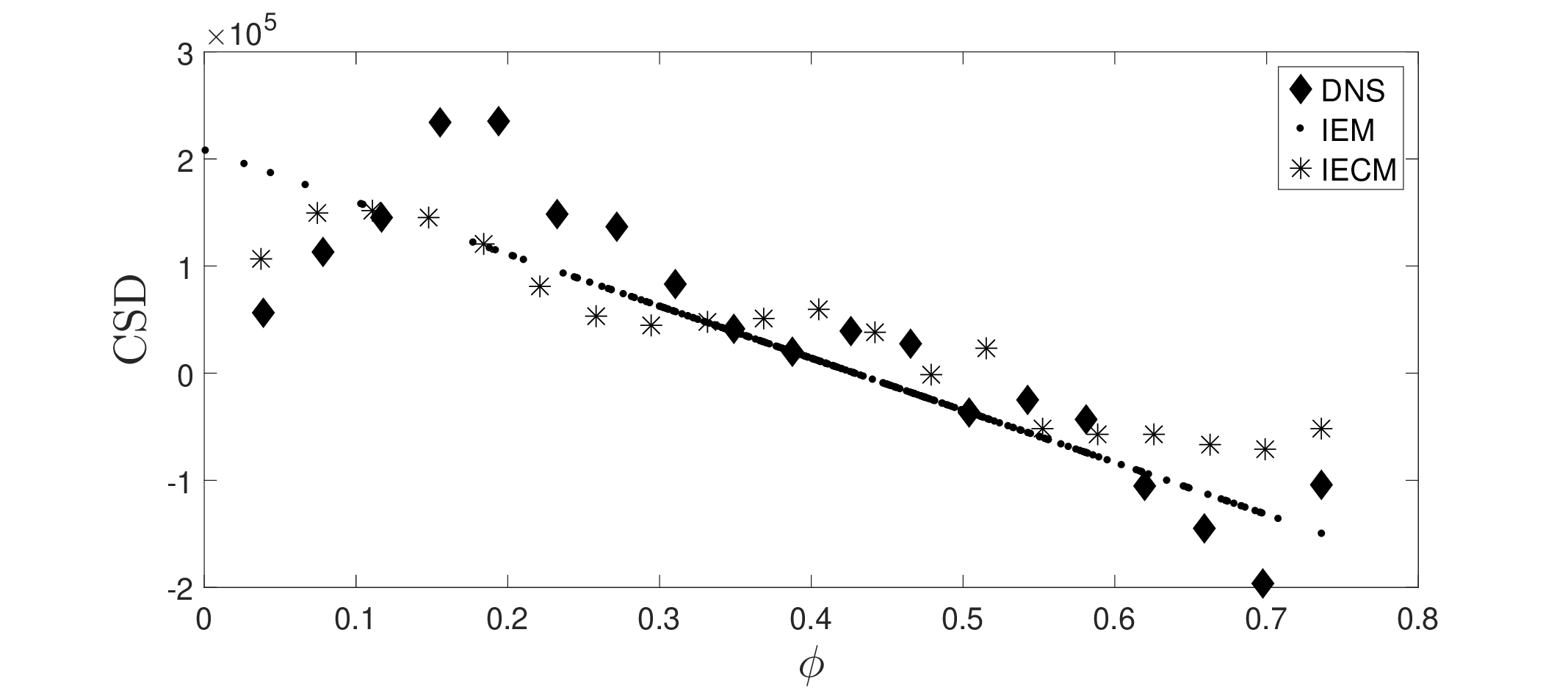}}}\hspace{5pt}
\caption{CSD $\big(\big\langle \frac{1}{\rho} \frac{\partial J_{\phi,j}}{\partial x_j} \big| \phi \big\rangle \big)$ modeled by the IECM model. CSD obtained from the DNS of heptane-air mixture at $P_0=100 atm$, $Re_0=1300$, $Le \neq 1$ with real gas equation of state and the generalized diffusion model. The mixing frequency for this model is same as that of the IEM model.}
\label{csdm100GNR1300_iecm}
\end{figure}

%------------------------------------
\begin{table}
\tbl{Comparison of calculated and corrected mixing frequencies ($\Omega_m$) for various test cases (non-reacting flows). $\Omega_{m,calculated}$ is determined from the definition in the IEM model, $\Omega_{m,corrected}$ is obtained from curve fitting method applied on the IEM model to fit the DNS data.}
{\begin{tabular}{lccccc} \toprule
Case & Species        & $P_0 (atm)$ & $Re_0$ & Diffusion   & $\frac{\Omega_{m,calculated}}{\Omega_{m,corrected}}$  \\ \midrule
1    & Heptane-Air    & 1           & 850    & Fickian     & 1.01                                             \\
2    & Heptane-Air    & 1           & 850    & Generalized & 1.20                                                        \\
3    & Heptane-Oxygen & 10          & 850    & Generalized & 12.11                                                        \\
4    & Heptane-Oxygen & 100         & 850    & Generalized & 11.07                                                      \\
5    & Heptane-Air    & 35          & 850    & Generalized & 23.87                                                    \\
6    & Heptane-Air    & 100         & 850    & Generalized & 11.07                                                    \\
7    & Heptane-Air    & 100         & 850    & Fickian     & 0.12                                                    \\
8    & Heptane-Air    & 35          & 1300   & Generalized & 10.96                                                  \\
9    & Heptane-Air    & 100         & 1300   & Generalized & 9.67                                                      \\
\bottomrule
\end{tabular}}
\label{mixfreqcomp}
\end{table}
%------------------------------------------------
The curve fits of the IECM model for the CSD are presented in Figs. \ref{csdm100GNR850_iecm}-\ref{csdm100GNR1300_iecm}. With the corrected mixing frequencies (from the IEM model), the trends predicted by the IECM model are qualitatively similar to those of the CSD. Although the curve fit is not exact, the norms of residuals of the model and the DNS data are comparable. 
% CSD BY IECM
%-------------------------------
%-------------------------------
\subsubsection{Modified Curl}
The model constant for Modified Curl model in Eq.(\ref{modcurl}) is determined to be $C_{MC}=1$ for flows at high pressures but for flow at atmospheric pressure with generalized diffusion the model constant is determined to be $C_{MC}=2$. Without the uniformly distributed random number $R_{ij}$, the MC model fits the CSD curve linearly. The term $R_{ij}$ causes interaction of random particles within an ensemble and produces good curve fits similar to CSD obtained from the DNS. However, random selection of particles from an ensemble for mixing is not physical as interaction is dependent on closeness of particles in physical and composition space. Thus, in turbulent flows with large fluctuations in density, temperature and scalars, the MC model may not accurately predict the CSD. The randomness factor added to the model introduces an artificial fluctuation in the trend of the model that may not be representative of the physical phenomenon. From Figs. \ref{csdm1FNI850}-\ref{csdm100GNR850_hepo} it is observed that the model follows the general trend of the CSD but tends to over-predict the DNS data. At atmospheric pressures the model predicts a trend similar to the DNS data but at higher pressures, the slope of model data is observed to be smaller than that of the CSD. 
%%%%%%%%%%%%%
\subsubsection{Mapping Closure}
\begin{figure}
\centering
{%
\resizebox*{12.5cm}{!}{\includegraphics{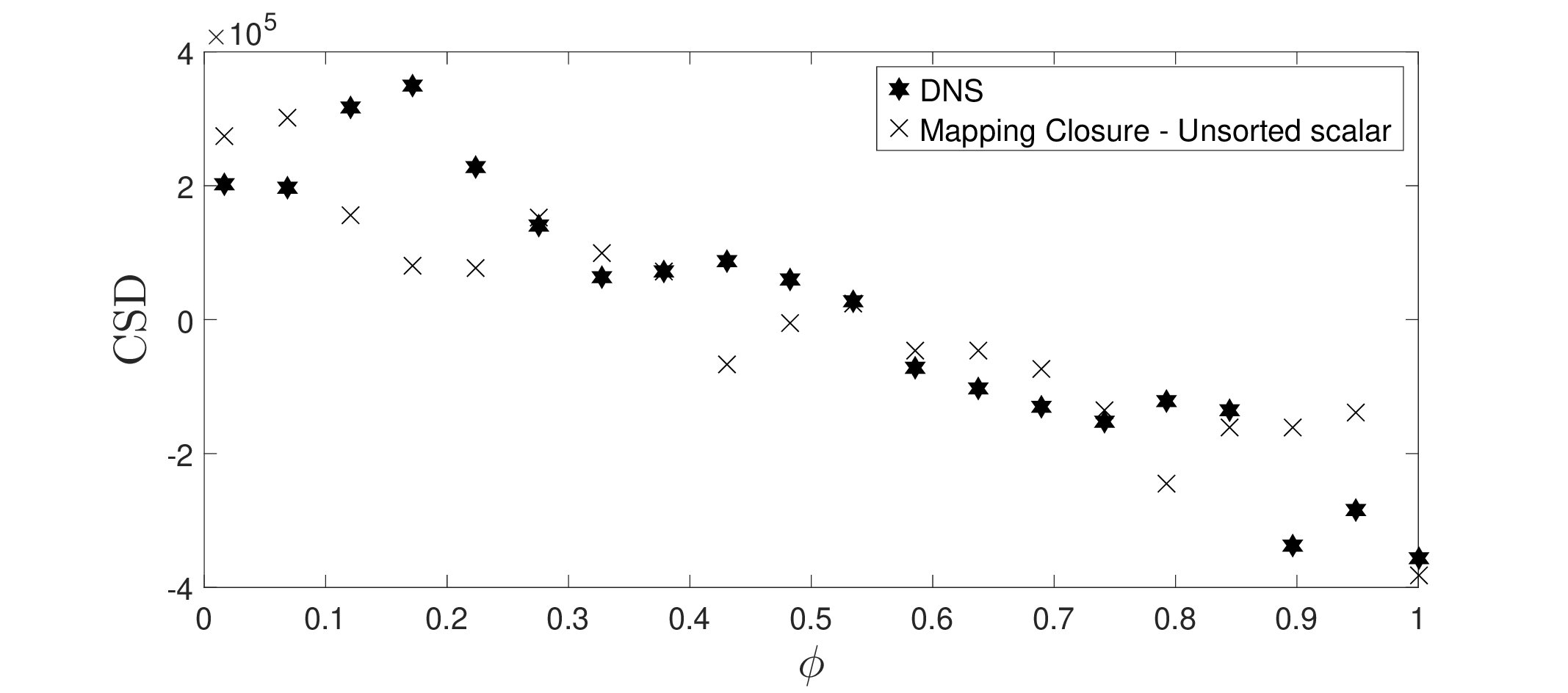}}}\hspace{5pt}
\caption{CSD $\big(\big\langle \frac{1}{\rho} \frac{\partial J_{\phi,j}}{\partial x_j} \big| \phi \big\rangle \big)$ modeled by the MAPPING model. The scalar value in the model remains unsorted. CSD obtained from the DNS of heptane-air mixture at $P_0=100 atm$, $Re_0=850$, $Le \neq 1$ with real gas equation of state and the generalized diffusion model.}
\label{csdm100GNR850_map_unsorted}
\end{figure}

The mapping closure (MAPPING) model presented in Section \ref{mapping} is conceptually an improvement over the IEM and MC models. However, in this model, the scalars within an ensemble are initially sorted and then mixed. This preserves the localness property of the model in composition space. In turbulent flows, particles with similar compositions tend to be local in physical space as well. The model constants ($C_{map}$) for various cases are presented in Table \ref{tab_modconst}. The curve fits of model in Figs. \ref{csdm1FNI850}-\ref{csdm100GNR850_hepo} suggest that the MAPPING model does not strictly follow the trend of the CSD. The slope of linear curve fit of the model data is much smaller than that of the CSD. This suggests that the model under-predicts the average conditional diffusion. However, at specific mixture fractions, the curve fit is representative of the actual phenomena. The small magnitude of slope of model data is a result of mixing of particles that are local within a composition space (due to sorted values of the scalar). However, if the scalars are unsorted, the trends predicted by the model are very similar to that of the CSD as observed in Fig. \ref{csdm100GNR850_map_unsorted}. The curve fits of the MAPPING model with unsorted scalar is similar to that of the Modified Curl model. Although the curve fit in this case is accurate, it is not physical. The idea of the MAPPING model for single scalar is extended to include multiple scalars in the EMST model. Evaluation of the EMST model is not included in this work.
%--------------MODEL CONSTANTS------------------
\begin{table}
\tbl{Model constants for the IEM, MC and MAPPING models for various test cases. $C_{IEM,fit}$, $C_{MC,fit}$ and $C_{MAP,fit}$ represent the factor by which the standard model constants were multiplied to fit the DNS data.}
{\begin{tabular}{ccccccccc} \toprule
Case & Species        & $P_0$ & $Re$ & EOS   & Diffusion   & $C_{IEM,fit}$ & $C_{MC,fit}$ & $C_{MAP,fit}$            \\ \midrule
1    & Heptane-Air    & 1     & 850  & Ideal & Fickian     & 1         & 1        & $-2 \times 10^{-2}$  \\
2    & Heptane-Air    & 1     & 850  & Real  & Generalized & 1         & 2        & $5 \times 10^{-2}$   \\
3    & Heptane-Oxygen & 10    & 850  & Real  & Generalized & 0.5       & 1        & $9 \times 10^{-3}$   \\
4    & Heptane-Oxygen & 100   & 850  & Real  & Generalized & 0.75      & 1        & $9.5 \times 10^{-3}$ \\
5    & Heptane-Air    & 35    & 850  & Real  & Generalized & 0.75      & 1        & $2 \times 10^{-2}$   \\
6    & Heptane-Air    & 100   & 850  & Real  & Generalized & 1         & 1        & $4 \times 10^{-2}$   \\
7    & Heptane-Air    & 100   & 850  & Ideal & Fickian     & 1         & 1        & $3 \times 10^{-3}$   \\
8    & Heptane-Air    & 35    & 1300 & Real  & Generalized & 1         & 0.875    & $2 \times 10^{-2}$   \\
9    & Heptane-Air    & 100   & 1300 & Real  & Generalized & 1         & 1        & $6 \times 10^{-2}$   \\
\bottomrule
\end{tabular}}
\label{tab_modconst}
\end{table}
%--------------------------------------
% UNSORTED MAPPING CLOSURE
%-----------3---------------
%-------------------------------
\section{Discussion}
\subsection{Mixing frequency}
The mixing frequency is a common factor in all molecular mixing models. It represents the rate of change of composition of particles within a subgrid volume. Although different models stem from different concepts of particle interaction, the rate of mixing in each model remains the same. Additional model constants may be used in different models to fit the data to experimental results. In each of the aforementioned models, the largest source of uncertainty (in curve fitting DNS data) is determined to be the mixing frequency. The form of mixing frequency used in this work may be expressed as,
\begin{equation} \label{eq_mixfreq}
\begin{split}
    \Omega_m = C_{\Omega} \frac{\Big \lbrack \gamma + \Big(\langle \rho \rangle_l \frac{C_R \Delta_G \sqrt{\xi}}{Sc_t} \Big) \Big \rbrack}{\langle \rho \rangle_l \Delta_G ^2}, \\
    \xi = |\langle u_i ^* \rangle_L \langle u_i ^* \rangle_L - \langle \langle u_i ^* \rangle_L \rangle_{l'} \langle \langle u_i ^* \rangle_L \rangle_{l'}|, \\
    u_i ^* = u_i - U_{ref,i}.
\end{split}
\end{equation}
This form is directly applicable only to flows at atmospheric pressures with the standard Fickian diffusion model. Even in the case of flow with the Fickian diffusion model at $P_0=100 atm$, the $\Omega_m$ calculated is off from the calibrated value by a factor of 10. This difference is more pronounced in flows that use a generalized diffusion model. The source of the difference is determined to be the diffusivity $\gamma$. The diffusivity is a constant in many applications of the Fickian diffusion model. At atmospheric pressures this assumption is often valid. However, at larger pressures it is more likely to have larger variations in diffusivities and cross diffusion effects become more important. Thus, a constant diffusivity assumption may lead to large errors in estimation of diffusion and mixing frequencies. By comparing the equations Eq.\ref{mass flux} and Eq.\ref{standard fickian}, a deviation factor for diffusivity, that takes into account the effects of cross-diffusion can be obtained. This can be expressed as,
\begin{equation}
    \gamma_{dev}= \frac{\Big( \frac{\partial J_j ^{\phi}}{\partial x_j} \Big)_{G}} {\frac{\partial ^2 \phi}{\partial x_i ^2}}.
\end{equation}
At atmospheric pressures, the deviation factor is almost unity ($\gamma_{dev} \approx \gamma$). However, at large pressures $\gamma_{dev} \neq 1$ since the standard Fickian diffusion model tends to over predict the diffusion. In this work, $\gamma_{dev}$ is used instead of $\gamma$ in Eq. (\ref{eq_mixfreq}) to calculate $\Omega_m$ for the models. A study on variation of parameters of mixing frequency ($\Omega_m = f(\gamma, \langle \rho \rangle_l, 1/ \Delta_G ^2, \sqrt{\xi})$) with ambient pressure is conducted. The factor that most significantly affects the mixing frequency is determined to be $\gamma$. For accurate modelling, an effective diffusion coefficient for the generalized diffusion model is required. A species-specific effective diffusion coefficient for a generalized diffusion model was derived by Ma \cite{mattdissertation} in which the generalized diffusion model for a species $\alpha$, ($\gamma ^{\alpha} _{j,G}$) is approximated to a model diffusion term of form $\gamma ^{\alpha} _{eff} \frac{\partial Y ^{\alpha}}{\partial x_j}$. A least square error method is then applied to the data to obtain an effective diffusion coefficient of the form,
\begin{equation}
    \gamma ^{\alpha} _{eff} = \frac{J^{\alpha} _{j,G} \frac{\partial Y^{\alpha}}{\partial x_j}}{\rho \frac{\partial Y^{\alpha}}{\partial x_j} \frac{\partial Y^{\alpha}}{\partial x_j}}.
\end{equation}
An effective Schmidt number for generalized diffusion models was also proposed,
\begin{equation}
    Sc ^{\alpha} _{eff} = \frac{\mu}{\rho \gamma ^{\alpha} _{eff}}.
\end{equation}
The readers are referred to the work of Ma \cite{mattdissertation} for a detailed analysis of molecular mixing models with the effective diffusion coefficient for generalized diffusion models at high pressures. 

The mixing frequency for turbulent flows can also be approximated by,
\begin{equation}
    \Omega_m = C_{k-\epsilon} \frac{\overline{\epsilon}}{\overline{k}}.
\end{equation}
In this equation, $\overline{\epsilon}$ and $\overline{k}$ represent the mean rate of dissipation and mean turbulent kinetic energy, respectively. In this work, the mean turbulent kinetic energy and dissipation rates are measured directly from the DNS at various pressures and the mixing frequencies are determined. These mixing frequencies are then used with the IEM model and the model constants $C_{k-\epsilon}$ are calibrated to fit the DNS data. The model constants for the corrected mixing frequencies are presented in Table \ref{tab_keps}. $C_{k-\epsilon}$ in most cases determined to be approximately 1. This suggests that, in LES with a generalized diffusion model, this form may be used as an alternate method to determine the mixing frequency.
\begin{equation}
\begin{split}
\Omega_m = \frac{\epsilon_{SGS}}{k_{SGS}}, \\
k_{SGS} = \frac{1}{2} u_i '' u_i '', \\
\epsilon_{SGS} = \nu \Big(\overline{\frac{\partial u_i}{\partial x_j} \frac{\partial u_i}{\partial x_j}} - \frac{\partial \overline{u_i}}{\partial x_j} \frac{\partial \overline{u_i}}{\partial x_j} \Big).
\end{split}
\end{equation}
%---------- C_{k-eps} --------------------
\begin{table}
\tbl{Model constants for mixing frequency obtained from mean turbulent kinetic energy and mean dissipation rates. $\overline{\epsilon}$ and $\overline{k}$ calculated from the DNS of mixing of heptane and oxygen.}
{\begin{tabular}{ccccccc} \toprule
$P_0 (atm)$ & $Re_0$ & EOS  & Diffusion   & $\overline{\epsilon}/\overline{k}$ & $\Omega_{m,corr}$  & $|C_{k-\epsilon}|$ \\ \midrule
1           & 850  & Real & Generalized & $-2.67 \times 10^3$  & $2.4 \times 10^3$  & 0.91               \\
35          & 850  & Real & Generalized & $-1.2 \times 10^5$   & $1.55 \times 10^5$ & 1.3                \\
100         & 850  & Real & Generalized & $-4.51 \times 10^5$  & $6.4 \times 10^5$  & 1.4                \\
\bottomrule
\end{tabular}}
\label{tab_keps}
\end{table}
%----------------------------------------
\section{Conclusion}
Results from direct numerical simulations of stationary jets at high pressure are presented. The formulation includes a real gas state equation, generalized diffusion and real temperature and pressure dependent property models. The results are used to test the pressure dependence of commonly used turbulence closure models. The statistics of the flow from the DNS at atmospheric pressures show good agreement with the results from various experimental and computational studies. The structural evolution of the flow at larger pressures can be assumed to be qualitatively similar. The DNS is used to extract the FMDF and the CSD at various pressures. The CSD term obtained from the DNS is compared against the CSD terms predicted by the mixing models. Although each of the mixing models discussed in this work are based on sound physical principles and are applicable to a large range of flows at atmospheric pressures, the CSD predicted by the models deviate from the CSD obtained from the DNS at large pressures. The cause for this deviation is determined to be the diffusivity term in the mixing frequency used by the models. At large pressures, the mixing frequency is under-predicted when a constant diffusivity (as in the case of standard Fickian diffusion model) is used and over-predicted when a deviation factor is used. The ideal solution is an effective diffusivity term that takes into account cross-diffusion effects at large pressures and can be used to estimate the mixing frequency. Derivation of such a term is beyond the scope of this work. Alternatively, the mixing frequency may be determined from the ratio of the mean dissipation rate and the mean turbulent kinetic energy of the flow.
\bibliographystyle{tfnlm}
\bibliography{interactnlmsample}
\end{document}